\documentclass{emulateapj}
\usepackage{graphics,graphicx,rotating}
\usepackage[colorlinks,allcolors=blue]{hyperref}
\usepackage[all]{hypcap} 
\usepackage{xspace}
\usepackage{amssymb}
\usepackage{amsmath}
\usepackage[percent]{overpic}
\usepackage{epstopdf}
\usepackage{subfigure}
\usepackage{framed}
\usepackage{xcolor}
\usepackage{natbib}
\begin{document}

\shorttitle{A fast rotating disky bulge in NGC 7025}
\title{High-resolution MEGARA IFU spectroscopy and structural analysis
  of a fast-rotating, disky bulge in NGC 7025}

\shortauthors{Dullo et al.}\author{Bililign T. Dullo$^{1}$}\author{ Mario Chamorro-Cazorla$^{1}$}\author{Armando Gil de
  Paz$^{1}$}\author{\'Africa Castillo-Morales$^{1}$}\author{Jesus  Gallego$^{1}$}
\author{Esperanza Carrasco$^{2}$} \author{Jorge  Iglesias-P\'aramo$^{3}$}\author{Raquel  Cedazo$^{4}$}
\author{Marisa L.\ Garc\'ia-Vargas$^{5}$}\author{Sergio Pascual$^{1}$}\author{Nicol\'as  Cardiel$^{1}$}
\author{Ana  P\'erez-Calpena$^{5}$}\author{\mbox{Pedro G\'omez-Cambronero}$^{5}$}\author{Ismael Mart\'inez-Delgado$^{5}$}\author{Cristina  Catal\'an-Torrecilla$^{1}$}
\affil{\altaffilmark{1} Departamento de F\'isica de la Tierra y
  Astrof\'isica, Universidad Complutense de Madrid, E-28040 Madrid,
  Spain; {\color{blue}bdullo@ucm.es}}
\affil{\altaffilmark{2} Instituto Nacional de Astrof\'isica, \'Optica
  y Electr\'onica, Luis Enrique Erro No.1, C.P. 72840, Tonantzintla,
  Puebla, Mexico}
\affil{\altaffilmark{3} Instituto de Astrof\'isica de Andaluc\'ia-CSIC,  Glorieta de
la Astronom\'ia s/n, 18008, Granada, Spain}
\affil{\altaffilmark{4} Universidad Polit\'ecnica de Madrid, Madrid, Spain}
 \affil{\altaffilmark{5} FRACTAL, S.L.N.E., Madrid, Spain}

\begin{abstract}

  Disky bulges in spiral galaxies are commonly thought to form out of
  disk materials (mainly) via bar driven secular processes, they are
  structurally and dynamically distinct from `classical bulges' built
  in violent merger events.  We use high-resolution GTC/MEGARA
  integral-field unit spectroscopic observations of the Sa galaxy NGC
  7025, obtained during the MEGARA commissioning run, together with
  detailed 1D and 2D decompositions of this galaxy's SDSS $i$-band
  data to investigate the formation of its disky (bulge) component
  which makes up $\sim 30\%$ of the total galaxy light. With a
  S\'ersic index $n \sim 1.80 \pm 0.24$, half-light radius
  $R_{\rm e} \sim 1.70 \pm 0.43$ kpc and stellar mass
  $M_{*} \sim (4.34 \pm 1.70) \times10^{10} M_{\sun}$, this bulge
  dominates the galaxy light distribution in the inner
  $R \sim 15\arcsec$ ($\sim 4.7$ kpc).  Measuring the spins
  ($\lambda$) and ellipticities ($\epsilon$) enclosed within nine
  different circular apertures with radii $R \le R_{\rm e}$, we show
  that the bulge, which exhibits a spin track of an outwardly rising
  $\lambda$ and $\epsilon$, is a fast rotator for all the apertures
  considered. Our findings suggest that this inner disky component is
  a pseudo-bulge, consistent with the stellar and dust spiral patterns
  seen in the galaxy down to the innermost regions but in contrast to
  the classical bulge interpretation favored in the past. We propose
  that a secular process involving the tightly wound stellar spiral
  arms of NGC 7025 may drive gas and stars out of the disk into the
  inner regions of the galaxy, building up the massive pseudo-bulge.

\end{abstract}

\keywords{
galaxies: bulges ---
galaxies: elliptical and lenticular, cD ---
galaxies: kinematics and dynamics ---
galaxies: photometry ---
galaxies: spiral ---
galaxies: structure
}

\section{Introduction}

A large number ($\sim$70\%) of observed spiral and S0 galaxies contain
a bulge component (e.g., \citealt[their Fig.~13]{2000A&A...361..863G};
\citealt{2003AJ....125.1073B}; \citealt{2006MNRAS.371....2A};
\citealt{2009ApJ...699..105C}). Excluding the innermost regions,
bulges are typically evident by their dominance of the inner stellar
light distribution of their host galaxies with respect to the outer
disk light profile. Earlier models of galaxy formation have predicted
that present-day bulges and elliptical galaxies are pressure-supported
systems with old stellar populations, described by the $R^{1/4}$ law
(\citealt{1948AnAp...11..247D}) and formed hierarchically through
mergers of smaller systems \citep[e.g.,][]{1988ApJ...331..699B,
  1996MNRAS.281..487K} or built in monolithic collapse of
protogalactic gas clouds \citep{1962ApJ...136..748E}. Confirming the
existence of such pressure-supported elliptical-like `classical
bulges' generally well described by the \citet{1968adga.book.....S}
$R^{1/n}$ model, subsequent studies have revealed disk-like
`pseudo-bulges\footnote{Throughout this paper, we use the term
  `pseudo-bulge' when referring to `disk-like' bulges, although this
  term is also sometimes used to refer to `boxy/peanut-shaped' bulges
  which are now understood to be thick bars seen edge-on
  \citep[e.g.,][]{1990A&A...233...82C, 1995ApJ...443L..13K,
    1999AJ....118..126B, 2005MNRAS.358.1477A, 2016ASSL..418...77L}.}'
which show significant rotational support, and tend to have range of
ages and low $n~(\la 2$) S\'ersic light profiles (e.g.,
\citealt{1982ApJ...257...75K}, \citealt{1982ApJ...256..460K};
\citealt{1993IAUS..153..387P}; \citealt{1993IAUS..153..209K};
\citealt{1996ApJ...457L..73C}; \citealt{1997AJ....114.2366C};
\citealt{2004ARA&A..42..603K}; \citealt{2005MNRAS.358.1477A};
\citealt{2007MNRAS.381..401L}; \citealt{2008AJ....136..773F};
\citealt{2008MNRAS.388.1708G}; \citealt{2009MNRAS.393.1531G}; 
\citealt{2013seg..book....1K}; \citealt{2015ApJ...799..226E};
\citealt{2016MNRAS.459.4109T}; \citealt{2017ApJ...848...87C}). The
difference between classical bulges and pseudo-bulges is thought to
reflect two distinct bulge formation paths, although both types of
bulges can coexist in a galaxy (\citealt{2003ApJ...597..929E};
\citealt{2005MNRAS.358.1477A}; \citealt{2008ASPC..390..232P};
\citealt{2015ApJ...799..226E}; \citealt{2016MNRAS.462.3800D}).

Pseudo-bulges are largely believed to be formed out of disks via
secular evolution driven by non-axisymmetric stellar structures such
as bars. However, little is known about the formation of pseudo-bulges
in unbarred disk galaxies, which account for roughly 30\% of local
disk galaxies (\citealt{1963ApJS....8...31D};
\citealt{2000ApJ...528..219K}; \citealt{2008ApJ...675.1141S};
\citealt{2015ApJS..217...32B};
\citealt{2018MNRAS.474.5372E}). Alternative pseudo-bulge formation
channels have recently been discussed in the literature.  For example,
pseudo-bulges built through gas-rich minor or/and major galaxy merger
events (e.g., \citealt{2005ApJ...622L...9S}
\citealt{2011A&A...533A.104E}; \citealt{2013ApJ...772...36G};
\citealt{2015A&A...579L...2Q}; \citealt{2016ApJ...821...90A};
\citealt{2018MNRAS.473.2521S}) or those that are formed at a high
redshift via starbursts \citep{2013MNRAS.428..718O}, or through clumps
that sink to the galaxy center by dynamical friction
\citep{2012MNRAS.422.1902I}.

Detailed structural and stellar kinematic studies of pseudo-bulges
enable us to discriminate between these different formation scenarios,
but the lack of robust bulge diagnostic criteria presents a major
challenge for the identification pseudo-bulges (e.g.,
\citealt{2013pss6.book...91G}; \citealt{2017A&A...604A..30N}).
\citet[his Section 4.3]{2013pss6.book...91G} provided cautionary
remarks about the misidentification of pseudo-bulges and classical
bulges when using criteria based e.g., on the bulge's S\'ersic index,
rotation and stellar age (\citealt[their
Section~4]{2004ARA&A..42..603K}; \citealt{2008AJ....136..773F}). A
related issue in the structural analysis of bulges is the failure to
account for small- and intermediate-scale components when modeling the
stellar light distributions of disk galaxies. \citet[see also
\citealt{2016ApJ...831..132G}]{2016MNRAS.462.3800D,2017MNRAS.471.2321D,2018MNRAS.475.4670D}
showed that neglecting to fit components such as bars, disks and
spiral arms, as a separate component, can systematically bias the
S\'ersic index and flux of the bulge component towards higher values
(e.g., \citealt{1996A&AS..118..557D}; \citealt{2005MNRAS.362.1319L};
\citealt{2008MNRAS.384..420G}).

Here we investigate the bulge of the spiral (Sa) galaxy NGC~7025 (RC3;
\citealt{1991rc3..book.....D}). This galaxy is a promising candidate
to explore the agents that drive the buildup of pseudo-bulges out of
disk materials via secular evolution in unbarred galaxies.  It is the
only unbarred isolated galaxy from a sample of 49 CALIFA galaxies by
\citet{2015MNRAS.451.4397H} which displays non-circular, bar-like
flows. The galaxy is among the first galaxies observed with the MEGARA
integral-field unit (IFU) as part of the commissioning run of the
MEGARA instrument.  The CALIFA kinematic maps for the galaxy by
\citet[their Figs.~A.1 and A.2]{2017A&A...597A..48F}, extracted using
the medium resolution CALIFA data, have a large field of view which
extends to the outer disk regions of the galaxy. These maps complement
our MEGARA data, which have higher spatial and spectral resolutions
and cover the inner half-light radius of the bulge. For comparison,
the medium resolution CALIFA data have a spectral resolution
$R_{\rm spec} \sim$ 1650 at $\sim 4500$
\AA~\citep{2017A&A...597A..48F}, while the lowest spectral resolution
of MEGARA is $R_{\rm spec} \sim 6000$.

Our isophotal analysis of the Sloan Digital Sky Survey (SDSS) images
of NGC~7025 reveals a `disky' bulge.  Recently,
\citet{2017A&A...604A..30N}, \citet{2018MNRAS.476.2137R} and
\citet{2018MNRAS.477..845G} fit a 2D S\'ersic bulge + exponential disk
model to the galaxy's SDSS images. \citet{2017A&A...604A..30N} argued
that the galaxy has a merger-built classical bulge with a
bulge-to-total flux ratio ($B/T) \sim 0.44$, consistent with the fit
by \citet{2018MNRAS.477..845G} which yielded a high S\'ersic index for
bulge ($n \sim 5.1$), suggestive of a classical bulge.  In contrast,
the bulge + disk fit by \citet{2018MNRAS.476.2137R} yielded a lower
value of $n$ $ (\sim 2.4)$ and a smaller $B/T$ of $ \sim 0.3$ for the
bulge component.

High-resolution structural and stellar kinematic studies of the bulges
of unbarred spiral galaxies such as NGC 7025 are of interest since
they can provide clues to the processes of galaxy formation and
evolution.  For the first time, we perform 1D and 2D, four component
(i.e., point source + bulge + intermediate-scale spiral-arm component
+ outer disk) fits to the SDSS data of NGC~7025. Combining these fits
together with the analyses of the galaxy images and our new
high-resolution MEGARA IFU spectroscopic observations, we favor a
pseudo-bulge interpretation for the bulge, which is built via secular
evolution driven by the spiral arms in the galaxy. Introducing the
MEGARA instrument in Section~\ref{Sec2.1.1}, we go on to describe our
MEGARA spectroscopic observations of the galaxy and the pertaining
data-reduction steps in Sections~\ref{Sec2.1.2} and \ref{Sec2.1.3},
respectively. The SDSS data for the galaxy and the corresponding data
reduction steps are detailed in Section~\ref{Sec2.2}.  Our structural
decompositions, a color profile and the analysis of the stellar
kinematics are given in Section~\ref{Sec3}.  We discuss NGC~7025's
pseudo-bulge formation and provide a brief summary in
Section~\ref{Sec4}.

Throughout this paper, we assume a cosmological
model with $H_{0} ~=~70$~km~s$^{-1}$~Mpc$^{-1}$,
$\Omega_{\Lambda} =0.7$ and $\Omega_{\rm m} =0.3$. This yields a
luminosity distance of 67.3 Mpc and a scale of 316 pc arcsec$^{-1}$
for NGC~7025 (NED\footnote{https://ned.ipac.caltech.edu}).

\begin{figure}
\hspace{-.3704976026340cm}   
\includegraphics[width=9.820066cm, height=4.cm]{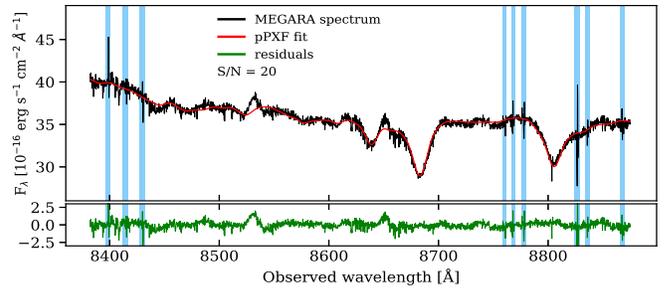}
 \caption{High-resolution MEGARA \mbox{HR-I} spectrum of NGC~7025, containing
  the Ca triplet absorption lines, extracted by
  co-adding all spaxels within the inner 1$\arcsec$ radius (black
  curve). The red and green curves show the {\sc pPXF} best fit to the
  spectrum and the associated residuals, respectively (see \ref{Sec3.2.1}). The blue vertical
  stripes are the mask of sky emission regions and absorption
  features omitted from the {\sc pPXF} fit. }
\label{Fig4} 
\end{figure}

\begin{figure}
\hspace{-.3704976026340cm}   
\includegraphics[width=9.820066cm, height=4.cm]{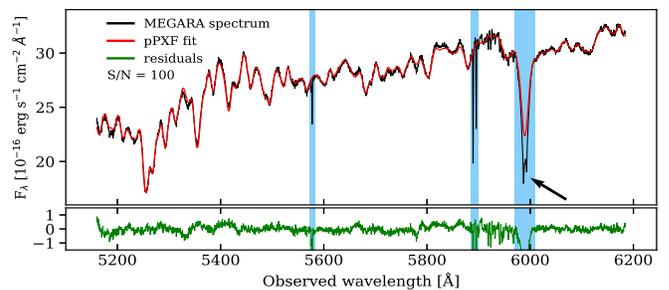}
\caption{Similar to Fig.~\ref{Fig4}, but now showing the MEGARA
  \mbox{LR-V} spectrum of NGC 7025 and the pertaining {\sc pPXF} fit
  together with the residuals about this fit. The arrow indicates the
  resolved interstellar \mbox{Na D} doublet. The blue vertical stripes
  show the mask of regions omitted from the {\sc pPXF} fit. }
\label{Fig4II} 
\end{figure}

\section{Data}
\subsection{MEGARA Spectroscopy}\label{Sec2.1}

\subsubsection{MEGARA}\label{Sec2.1.1}

MEGARA\footnote{https://guaix.fis.ucm.es/megara}
(Multi-Espectr\'ografo en GTC de Alta Resoluci\'on para Astronom\'ia)
is a new optical integral field unit (IFU) and multi-object (MO)
spectrograph installed on the 10.4-m Gran Telescopio CANARIAS (GTC) in
La Palma (Gil de Paz et al.\ 2018, in prep.). Both the MEGARA IFU and
MOS have low, medium and high spectral resolutions $R_{\rm spec}$ of
$\sim$ 6000, 12,000 and 20,000, respectively. The MEGARA IFU/MOS VPH
gratings cover a wavelength range of $\sim$ 3653$-$9686 {\AA}.

The MEGARA IFU encompasses 567 contiguous hexagonal fibers, each with
a long diagonal of $0\farcs62$, resulting in an $\sim$ 12.5 $\times$
11.3 arcsec$^{2}$ field of view in the shape of a rectangle. A total
of 8 static fiber bundles with 56 dedicated hexagonal fibers,
located at the outermost parts of the field of view far from the IFU
($1\farcm5 \la R \la 2\arcmin$), deliver simultaneous sky
observations. 

\subsubsection{MEGARA observations}\label{Sec2.1.2}

MEGARA IFU spectra of NGC~7025 were obtained in all VPH gratings as
part of the instrument's commissioning run between 2017 June 24 and
August 31. In this paper, we focus only on the high-resolution and
low-resolution spectra of the galaxy obtained in the MEGARA's
VPH863-HR grating (henceforth referred to as HR-I) and VPH570-LR
grating (henceforth LR-V), respectively (Figs.~\ref{Fig4} and
~\ref{Fig4II}).  The HR-I spectra have a wavelength range
\mbox{$ \sim$ 8372$-$8882 {\AA}}, while the LR-V spectra cover $ \sim$
\mbox{5144 $-$6168 {\AA}}. This yields reciprocal linear
dispersions of $0.13 $ {\AA} pixel$^{-1}$ and $0.27 $ {\AA}
pixel$^{-1}$ for the \mbox{HR-I} and \mbox{LR-V} spectra,
respectively. For the \mbox{HR-I} spectra, the full width at
half-maximum (FWHM) resolution at the central wavelength is 0.42 {\AA}
$\approx$ 15 km~s$^{-1}$, while for the \mbox{LR-V}, FWHM $\sim$ 0.95
{\AA} $\approx$ 50 km~s$^{-1}$ (\citealt{2016SPIE.9908E..1KG}).

The LR-V and HR-I spectra of NGC~7025 were obtained on the nights of
2017 August 1 and 2, respectively.  We obtained three exposures of NGC
7025 per VPH: 3$\times$900 s both for the \mbox{HR-I} and \mbox{LR-V}
VPH gratings. The seeing conditions ranged between
0$\farcs7 -1\farcs1$. To flux calibrate the \mbox{HR-I} spectra,
3$\times$45 s exposures of the flux standard star \mbox{BD+174708}
were obtained under similar seeing and airmass conditions to those of
the science exposures. For the \mbox{LR-V} spectra, we obtained
\mbox {3$\times$30 s} exposures of the standard star \mbox{BD+332642}.  We
also obtained calibration data including Th-Ne and Th-Ar arc-lamps,
halogen lamps and twilight spectra.

\begin{figure}
\includegraphics[angle=270,scale=0.6]{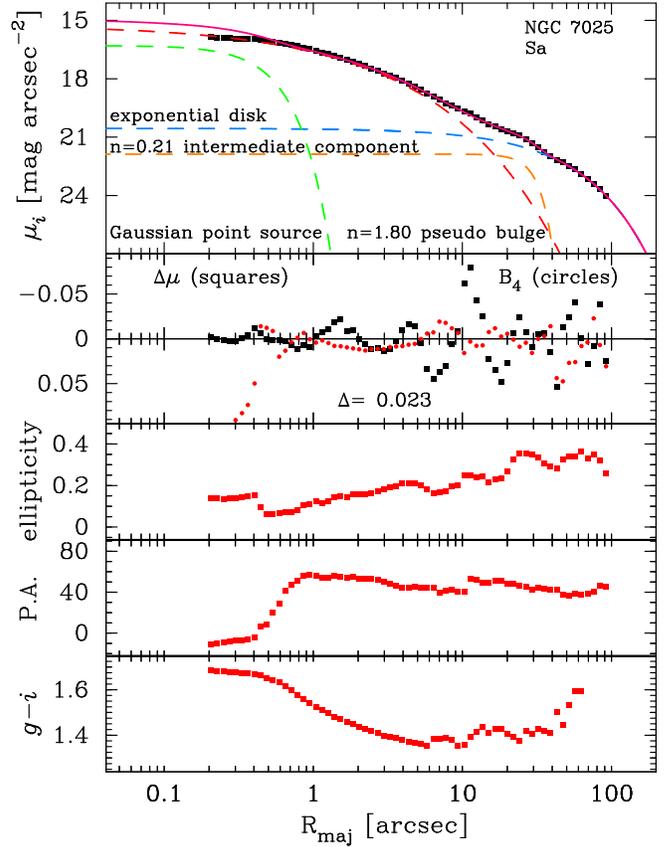}
\caption{Major-axis SDSS $i$-band surface brightness ($\mu_{i}$),
  isophote shape parameter (B$_{4}$), ellipticity ($\epsilon$),
  position angle (P.A., measured in degrees from north to east), and
  color profiles of NGC 7025.  Note that the x-axis is on a
  logarithmic scale. The typical errors associated with B$_{4}$,
  $\epsilon$ and P.A. are 20$-$30\%, 4\% and 5\%, respectively.  The
  errors on $\mu_{i}$ and the $g-i$ color are typically 0.015 mag
  arcsec$^{-2}$ and 0.02 mag. The dashed curves show the
  four-component decomposition of the major-axis light profile, i.e.,
  inner point source (green curve) + pseudo-bulge (red curve) +
  intermediate spiral-arm component (orange curve) + outer disk (blue
  curve). The solid curve (magenta) shows the complete fit with a
  small rms residual $\Delta$.}
\label{Fig1} 
 \end{figure}

\subsubsection{Data reduction}\label{Sec2.1.3}

The raw IFU spectra of NGC 7025, `cleaned' of cosmic rays
interactively using the {\sc
  cleanest}\footnote{http://cleanest.readthedocs.io/en/latest/}
(Cardiel et al.\ 2018, in prep.) software package, were processed
using the MEGARA data reduction pipeline
MDRP\footnote{http://megara-drp.readthedocs.io/en/latest/} (S. Pascual
2018, in prep.). First, we masked the bad pixels in the IFU and
subtracted a bias frame.  The spectra were then processed using the
MDRP tasks {\sc trace} and {\sc modelmap} by tracing the fibers across
the flat halogen lamp frames. MDRP was used to perform the wavelength
calibration of the HR-I and LR-V spectra using ThNe arc-lamp and ThAr
arc-lamp frames, respectively, to an accuracy of $\la$ 0.01 \AA ~ and
$\la$ 0.03 \AA ~root-mean-square values. MDRP corrects for
spaxel-to-spaxel sensitivity and fiber-to-fiber transmission
variations using our halogen and twilight flat frames,
respectively. The spectra were flux calibrated by comparing our
observations of the flux standard star with the star's calibrated
spectra provided by the CALSPEC calibration data
base\footnote{http://www.stsci.edu/hst/observatory/crds/calspec.html}. Finally,
MDRP subtracts a median sky spectrum, determined using spectra from
the 56 dedicated sky fibers, to generate the fully reduced row-stacked
spectra (RSS), where the rows pertain to the spectra of the individual
623(=567+56) science and sky fibers. The MDRP task {\sc cube}
transforms these RSS files into data cubes, but we use the RSS data
for the stellar kinematic study in this paper. Figs.~\ref{Fig4} and
\ref{Fig4II} show the reduced central ($R \la 1\arcsec$) spectra of
NGC 7025.

\begin{figure*}
\begin{center} 
\hspace*{-1.79cm}   
\includegraphics[angle=270,scale=0.69899905]{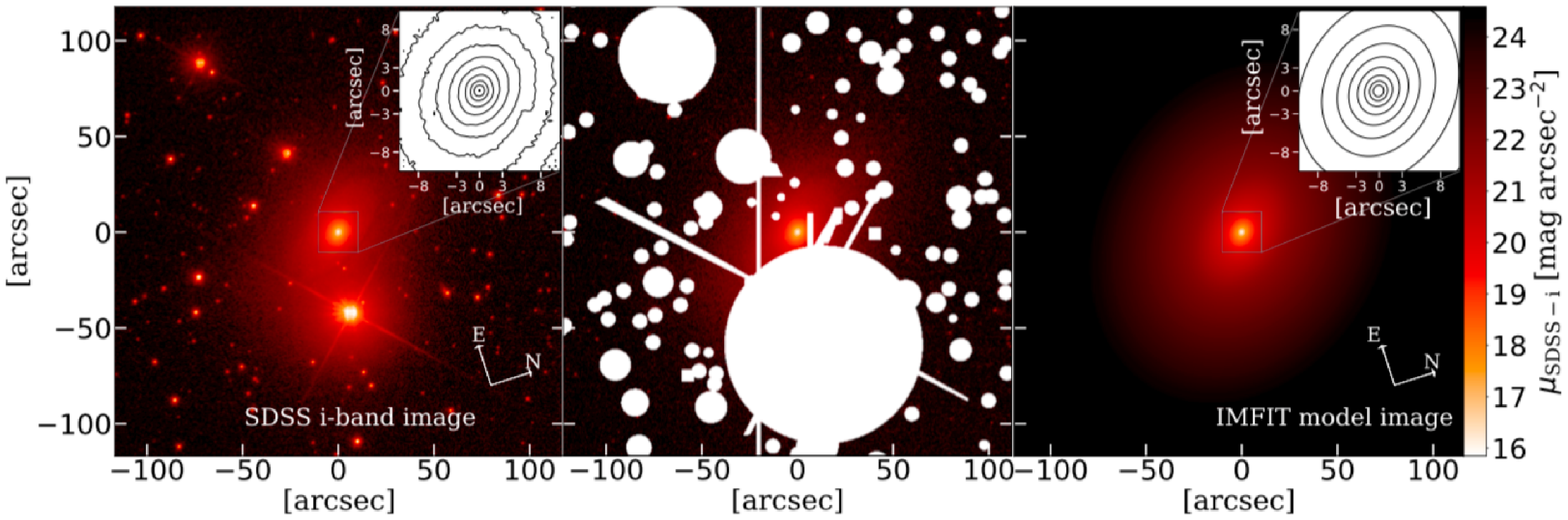} 
\vspace*{-4.70cm}   
\label{Fig2} 
\caption{ Left-hand panel:  SDSS $i$-band image of NGC 7025. Middle
  panel: masked regions (i.e., white areas). Right-hand panel: {\sc imfit} model
  image of NGC~7025. The insets show the surface brightness contours. }
\end{center}
\end{figure*}

\subsection{SDSS imaging data}\label{Sec2.2}

There are no high-resolution {\it Hubble Space Telescope (HST)} images
available for NGC~7025. Therefore, $13\farcm51\times9\farcm83$, bias
subtracted and flat-fielded SDSS $g$- and $i$-band images of NGC~7025
were retrieved from the SDSS Data Release 7 (DR7) data
base\footnote{http://www.sdss.org}. These images
contain a `soft bias' of 1000 DN, added to each pixel. In order to
better correct for the galaxy's nuclear dust spiral, we used the
$i$-band image for the detailed structural analysis
(see Figs.~\ref{Fig1} and \ref{Fig2}). We used the $g$- and $i$-band data to extract the
$g- i$ color profile for the galaxy (Fig.~\ref{Fig1}).

\begin{center}
\begin{table*} 
\setlength{\tabcolsep}{0.037239948in}
\caption{Structural parameters}
\label{Table1}
\begin{tabular}{@{}llccccccccccccccc@{}} 
\hline
&&&&&&&1D fit (major-axis)\\
\hline
$ \mu_{\rm e,bul} $ & $R _{\rm e,bul}$ &$R _{\rm e,bul}$ & $n
                                                                 _{\rm
                                                                 bul}$&$\mu_{\rm
                                                                          e,int}
                                                                          $
                                                               & $R
                                                                 _{\rm
                                                                 e,int}$&$n_{\rm,
                                                                          int}$&$
                                                                                 \mu_{\rm
                                                                                 0,disk}
                                                                                 $&$h$&$m_{\rm
                                                                                        pt}$&$M_{
                                                                                              i,\rm
                                                                                              bul}$
                                                                                             &$B/T$&$B/D$&$M/L_{i,\rm
                                                                                                           bul}$&log
                                                                                                                  ($M_{*,\rm
                                                                                                                          bul}/M_{\sun}$)&\\
&(arcsec)&(kpc)&&&(arcsec)&&&(arcsec)&(mag)&(mag)&&\\
(1)&(2)&(3)&(4)&(5)&(6)&(7)&(8)&(9)&(10)&(11)&(12)&(13)&(14)&(15)&\\
\multicolumn{1}{c}{} \\              
\hline       
                 18.77 &  5.23 &1.65 & 1.80  &22.04&20.15 &  0.21    &
                                                                      20.55
                                                                          &28.89&
                                                                                   16.95&-22.53&0.30 &0.48&0.63&10.62
  \\
  
&&&&&&&2D fit ({\sc imfit})\\
\hline       
                 18.61& 4.68 &1.48 & 1.52 & 21.82&17.90 &  0.24   &
                                                                        20.44
                                                                          &23.76&
                                                                                    16.01
                                                                                  & -22.59&0.28 &0.48&0.63&
                                                                                                          10.65\\
\hline
&20\%&&15\%&&25\%&20\%&&15\%&&&15\%&15\%&15\%&40\%&\\
\hline
\end{tabular} 
Notes.--- 1D and 2D structural parameters from Gaussian point source +
S\'ersic pseudo-bulge + S\'ersic intermediate-component + exponential
outer disk model fit to the SDSS $i$-band major-axis light profile and
to the SDSS $i$-band image of NGC 7025, respectively
(Figs.~\ref{Fig1}, \ref{Fig2} and \ref{Fig3}).  Cols.\ $1- 7$:
S\'ersic model parameters for the pseudo-bulge and intermediate-scale
component. Cols.\ 8$-$9 best-fitting parameters for the outer
exponential disk. Col.\ 10: apparent magnitude of the point source.
Cols.\ $11$: absolute magnitude of the pseudo-bulge  calculated
using cols.\ $1- 3$ and corrected for Galactic extinction, surface
brightness dimming and internal dust attenuation. Cols.\ $12-13$:
bulge-to-total and bulge-to-disk flux ratios. The above dust and
surface brightness dimming corrections were not
applied to these flux ratios. Cols.\ $14-15$:
$i$-band mass-to-light ratio ($M/L_{i,\rm bul}$) and stellar mass of
the pseudo-bulge ($M_{*,\rm bul}$).  $\mu_{\rm e}$ and $\mu_{\rm 0}$ are in mag arcsec$^{-2}$. The last row shows the
1$\sigma$ uncertainties on the fit parameters, flux ratios and stellar
masses, see the text for details. The uncertainties associated with $m_{\rm pt}$ and $M_{\rm i,bul}$ are
$0.95$ mag and $0.2$ mag, respectively. The uncertainty on $\mu_{\rm e}$ and $\mu_{\rm 0}$ is
$\sim$ 0.025 mag arcsec$^{-2}$.
\end{table*}
\end{center}

\subsubsection{Data reduction}\label{Sec2.2.1}

The SDSS images were reduced using standard {\sc iraf} tasks
\citep[references therein]{2017MNRAS.471.2321D, 2018MNRAS.475.4670D}.
Subtracting the `soft bias', we determined the sky background levels
as the average of the median of the sky values from several 10
$\times$ 10 pixel boxes, far from the galaxy. An initial mask was
generated by running SEXTRACTOR \citep{1996A&AS..117..393B}, which was
then combined with a manual mask to avoid the dust spiral in the
galaxy as well as the bright foreground stars, background galaxies,
and chip defects in the images (Figs.~\ref{Fig2} and ~\ref{Fig3}). We
used this mask and fit the {\sc iraf} task {\sc ellipse}
\citep{1987MNRAS.226..747J} to the sky subtracted images following the
steps outlined in \citet{2017MNRAS.471.2321D, 2018MNRAS.475.4670D}.
The results of the {\sc ellipse} fitting are fed into the IRAF task
{\sc bmodel} to create a model image of the galaxy.  This model image
was subtracted from the science image to create an initial residual
image and an improved mask (Fig.~\ref{Fig2}, middle).  We then re-ran
the {\sc ellipse} fitting and {\sc bmodel}. Fig.~\ref{Fig1} shows
results of the {\sc ellipse} fitting including the major-axis surface
brightness, isophote shape parameter ($B_{4}$), ellipticity and
position angle profiles of NGC 7025. Fig.~\ref{Fig3} (left) shows the
final residual image. The surface brightnesses were calibrated using
the zero-points, extinction coefficients and airmass values given in
the DR7 tsField files.

\begin{center}
\begin{table} 
\setlength{\tabcolsep}{0.03239948in}
\caption{Structural parameters}
\label{Table2}
\begin{tabular}{@{}llccccccccccccccc@{}}
\hline
&&&2D fit ({\sc imfit})\\
\hline
P.A.$_{\rm bul}$ & $\epsilon_{\rm bul}$ &c$_{0,\rm bul}$&P.A.$_{\rm
                                                                          int}
                                                                          $
                                                               & $\epsilon_{\rm int}$&c$_{0,\rm int}$&P.A.$_{\rm disk}$ &  $\epsilon_{\rm disk}$ 
                                                                                 \\
($^{\circ}$)&&&($^{\circ}$)&&&($^{\circ}$)&\\
(1)&(2)&(3)&(4)&(5)&(6)&(7)&(8)&\\
\multicolumn{1}{c}{} \\              
\hline       
                 46.78 &  0.21&-0.05 & 40.01  &0.37&-0.01& 42.41   &  0.25    \\                                                       
\hline
10\%&10\%&10\%&10\%&10\%&10\%&10\%&20\%&\\
\hline
\end{tabular} 
Notes.---   Similar to Table~\ref{Table1} but showing the remaining structural
parameters from the 2D decomposition with {\sc imfit}. The
{\sc imift} isophote shape parameter c$_{0}$ as given by \citet[their
Equation 25]{2015ApJ...799..226E} is positive/negative for boxy/disky
isophotes, not to be confused with that of the IRAF/{\sc ellipse} $B_{4}$.\\\\
\end{table}
\end{center}

\begin{figure*}
\begin{center}
\setlength{\tabcolsep}{0.0348in}
\begin{tabular}{@{}llcccc@{}}
\multicolumn{1}{c}{} \\       
\hspace*{-1.4390700052250772599cm}   
\includegraphics[angle=270,scale=0.71595]{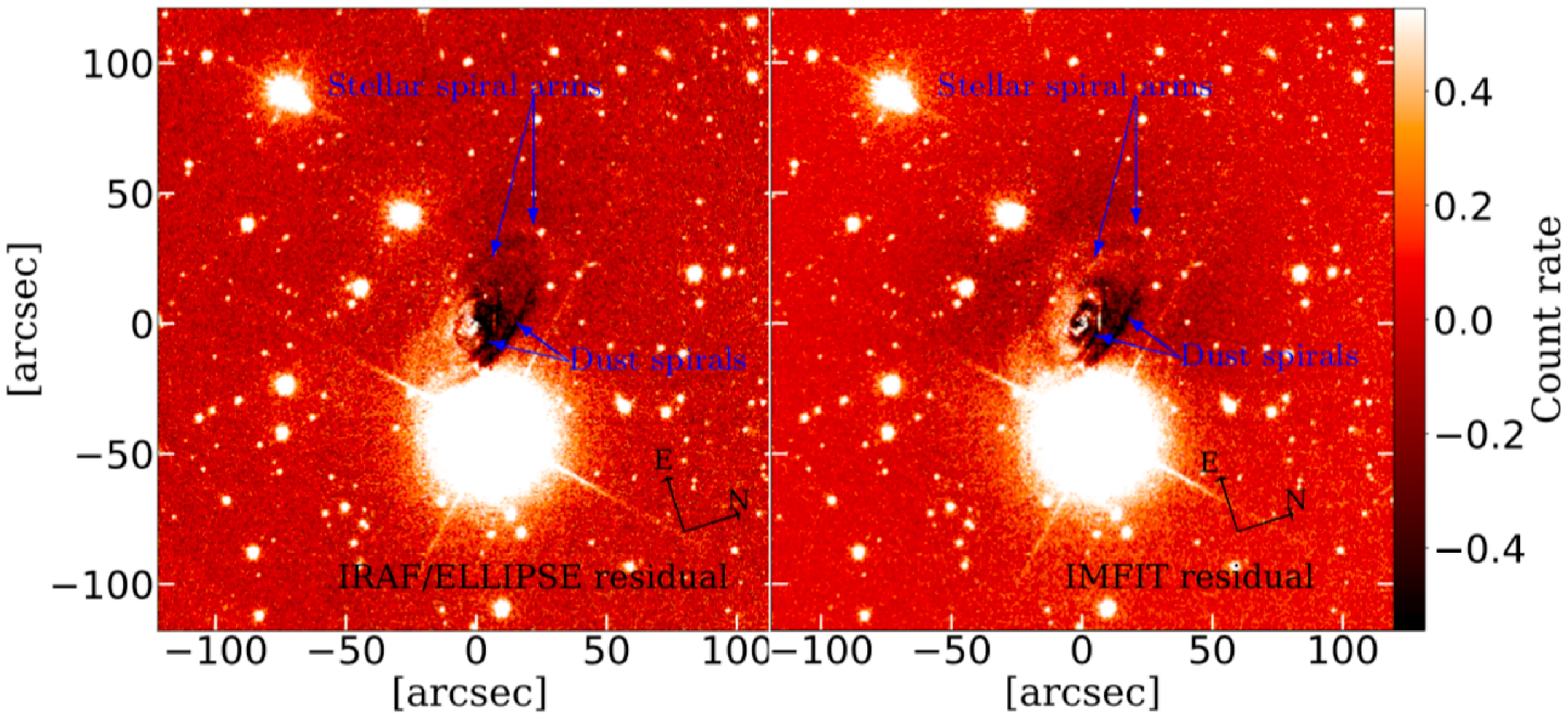}
\end{tabular} 
\label{Fig3} 
\vspace*{-4.0800772599cm}   
\caption{Residual images which are generated  after
  subtracting the {\sc iraf/bmodel} and {\sc imfit} model images from
  the original SDSS image of NGC~7025 reveal dust spiral patterns and two
  tightly wound stellar spiral arms. }
\end{center}

\end{figure*}



\begin{table*} 
\caption{Structural parameters}
\label{Table22}
\begin{tabular}{@{}llccccccc@{}}
\hline
&&&&2D fit ({\sc galfit})\\
\hline
Component&S\'ersic&$m_{i}$ (mag)&$n$&$R_{\rm
                                                e}$ ($\arcsec$)&P.A. ($^{\circ}$) & $\epsilon$ &c$_{0}$&$M_{i}
                                                                          $
  (mag)\\
&log&$R_{\rm in}$ ($\arcsec$)&$R_{\rm out}$ ($\arcsec$)&$\theta_{\rm rot}$
                                 ($^{\circ}$)&$R_{\rm ws}$ ($\arcsec$)&
                                                             $\theta_{\rm
                                                             incl}$
                                                             ($^{\circ}$)&$\theta_{\rm
                                                                           sky}$
                                                                           ($^{\circ}$)&
  \\
                                                                       
&fourier & mode: amplitude & phase  ($^{\circ}$)& mode: amplitude& phase  ($^{\circ}$)
                                                              
                                                                                 \\
(1)&(2)&(3)&(4)&(5)&(6)&(7)&(8)&\\
\multicolumn{1}{c}{} \\              
\hline       
 Bulge&   S\'ersic&  12.40 &  1.92&5.21 & 45.49 
                                              &0.20&-0.01& -22.55
                                                                     
  \\        
  
Intermediate spiral-arm&   S\'ersic&   14.26 & 0.45& 16.22 & 47.79
                                         &0.42&---&---  
  \\  
 component    & log&  18.61 &5.54&80.8  &2.61&52.69& 54.82 &---   \\                                                      
&   fourier&      1:0.40&0.22 &  2:0.36&9.15& ---&---  &---    \\        
\hline
\hline
\end{tabular} 

Notes.---   Similar to Tables~\ref{Table1} and  \ref{Table2} but
showing here the bulge and spiral-arm structural
parameters from the 2D decomposition of the SDSS $i$-band image of
NGC~7025 with {\sc galfit}
\citep[their equation 29]{2010AJ....139.2097P}. The {\sc galfit} best-fitting
parameters for the point source and outer disk are similar to those
from  {\sc imfit}. We fit the {\sc galfit} logarithmic
spiral (log-tanh) function to the intermediate spiral-arm component,
see the text for details. $m_{i,\rm {bul}}$ is apparent magnitude of the
bulge returned by  {\sc galfit}.  $M_{i,\rm {bul}}$ absolute magnitude of the
bulge calculated using $m_{i,\rm {bul}}$, corrected for Galactic
extinction, surface brightness dimming and internal dust
attenuation. $R_{\rm in}$ and $R_{\rm out}$ are the inner
and outer radii of the spiral with cumulative rotation of $\theta_{\rm
  rot}$  and  winding scale radius of $R_{\rm ws}$. $\theta_{\rm
  incl}$ and $\theta_{\rm
  sky}$
are
the
spiral's
sky
inclination
and 
position
angles,
respectively. Errors
on
the
S\'ersic fit
parameters
are as
in
Tables~\ref{Table1}
and
\ref{Table2}. 
\end{table*}

\section{Results}\label{Sec3}

\subsection{Photometric structural analysis}\label{Sec3.1}

Fig.~\ref{Fig1} shows our one-dimensional (1D) four-component
decomposition of the SDSS $i$-band surface brightness profile of
NGC~7025.  The fit, which contains a Gaussian ($n=0.5$) nuclear
component, a S\'ersic bulge, a S\'ersic intermediate-scale component
and an outer exponential ($n=1$) disk, describes the light profile
very well as revealed by the small root-mean-square (rms) residual
value of $\sim 0.023$ mag arcsec$^{-2}$ (Fig.~\ref{Fig1}). This
detailed multi-component fit was performed following the similar
procedures as those in \citet{ 2017MNRAS.471.2321D,
  2018MNRAS.475.4670D}. For each component, the model profile was
convolved with a Gaussian PSF with FWHM $\sim $0$\farcs98$. The FWHM
of the PSF, which we measured using several stars in the SDSS $i$-band
image, agrees with that given in the SDSS DR7 tsField file.

We additionally performed a 2D, four-component decomposition of the
SDSS $i$-band image using {\sc
  imfit}\footnote{\url{http://www.mpe.mpg.de/~erwin/code/imfit/}}. We
used the result from the 1D fitting as an input for the 2D fitting.  To
convolve the model images, we used a Moffat PSF image generated using
the {\sc imfit} task {\sc makeimage}. Fig.~\ref{Fig2} (right) shows
the best-fitting 2D Gaussian nuclear component + S\'ersic bulge +
S\'ersic intermediate-scale component + outer exponential disk model
image that describes the 2D light distribution of the galaxy. {\sc
  imfit} has subtracted this model image from the galaxy image to
produce the residual image (Fig.~\ref{Fig3},
right). Tables~\ref{Table1} and \ref{Table2} list the best-fitting
model parameters from the 1D and 2D decompositions.

Both the {\sc ellipse} and {\sc imfit} residual images reveal a dust
spiral pattern which extends from the galaxy center out to $R \sim24\arcsec$,
indicative of a recent star formation activity in the galaxy. Also
apparent in these residual images are two tightly wound stellar
spiral arms over the regions where the intermediate-scale component
dominates the bulge light. This S\'ersic intermediate-scale
(spiral-arm) component with S\'ersic index $n\sim 0.22$, dominated by
the outer disk at all radii, is also evidenced by the disky isophotes
(i.e., the {\sc ellipse} $B_{4} > 0$ and {\sc imfit} C$_{0} < 0$) and
high ellipticity ($\epsilon \sim 0.35$) at
$22\farcs5 \la R \la 32\arcsec$. As can be seen in Fig.~\ref{Fig1},
the ellipticities and orientations of the (spiral-arm) component and
outer disk from the {\sc ellipse} isophote fitting are,
unsurprisingly, similar.

\begin{figure*}
\begin{center}
\begin{tabular}{@{}llcccc@{}}
\multicolumn{1}{c}{} \\       
\hspace*{-.883700052250772599cm}   
\includegraphics[angle=270,scale=0.6595]{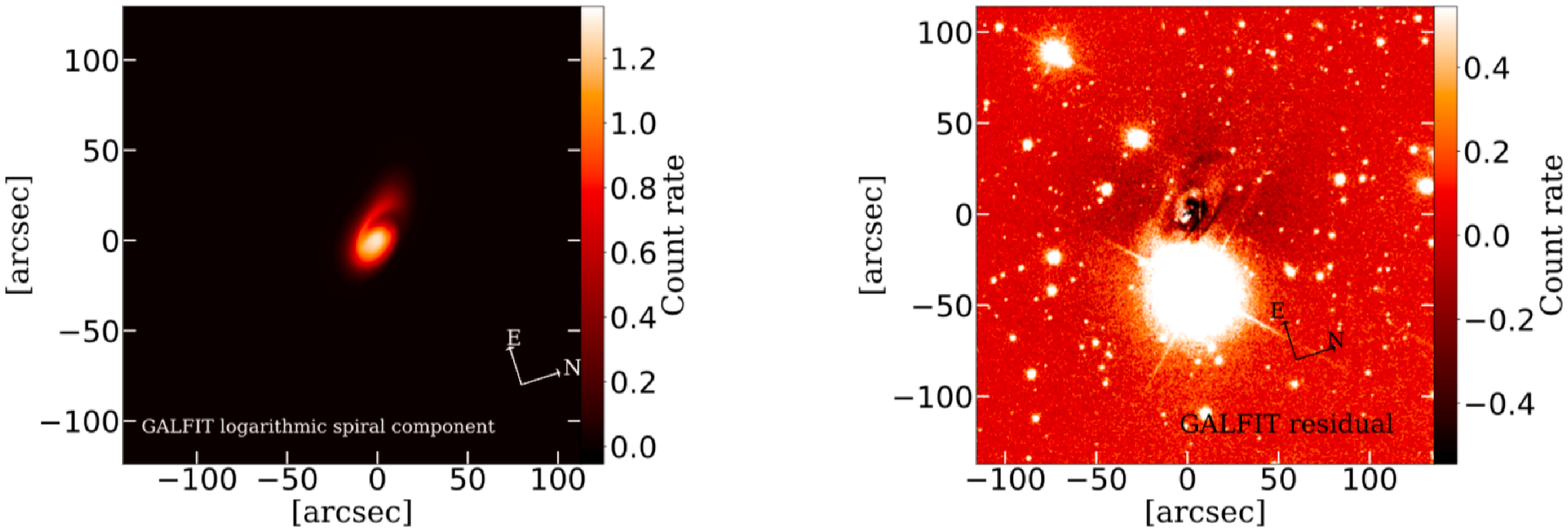}
\end{tabular} 
\label{Fig333} 
\vspace*{-3.800772599cm}   
\caption{Left-hand panel: {\sc galfit} model image of the spiral-arm
  component in NGC~7025. Right-hand panel: residual image which is
  generated after subtracting the {\sc galfit} model image of
  NGC~7025, which consists the spiral-arm component (left panel), from
  the original SDSS image of the galaxy. }
\end{center}
\end{figure*}

Given that we fit a S\'ersic model to the \mbox{intermediate-scale}
\mbox{ spiral-arm} component, the flux and S\'ersic index of the bulge
as well as the flux of outer disk may be biased to lower
values. Concerned about this, we went further and performed a 2D,
four-component decomposition of the galaxy image using {\sc galfit}
\citep{2010AJ....139.2097P}, see Table~\ref{Table22} and
Fig.~\ref{Fig333}, fitting the {\sc galfit} logarithmic spiral
(log-tanh) model image (Fig.~\ref{Fig333}, left, see \citealt[their
equation 29]{2010AJ....139.2097P}) to the spiral-arm component. The
{\sc galfit} residual image in Fig.~\ref{Fig333} (right) shows some
stellar spiral-arm structures that we missed in our modeling. However,
we avoid performing a more sophisticated fit due to strong
degeneracies between the fit parameters. Our {\sc galfit} and {\sc
  imfit} decompositions agree very well except for the \mbox
{spiral-arm} component model. The {\sc galfit} best-fitting parameters
for the point source and outer disk are similar to those from {\sc
  imfit}, thus we do not show them in Table~\ref{Table22}. The
best-fitting parameters of the bulge (e.g., $n$, $R_{\rm e}$ and
$M_{i}$) from {\sc galfit} are in good agreement with those from {\sc
  imfit} (see Tables~\ref{Table1}, \ref{Table2} and
\ref{Table22}). For example, $n_{\rm bul,\text{\sc {galfit}}}$
($\sim$1.92 $\pm~ 0.30$) agrees well within the uncertainties with
$n_{\rm bul,\text{\sc {imfit}}}$ ($\sim$1.52 $\pm ~0.23$).

The agreement between our 1D and 2D decompositions is very good
(Fig~\ref{Fig3}, Tables~\ref{Table1} and \ref{Table2}), although the
apparent magnitude of the innermost nuclear component, obtained using the
best-fitting 1D (major-axis) structural parameters and the pertaining
ellipticity ($\sim17.0$ mag), differs somewhat from that of the 2D fit
($\sim16.0$ mag).  Since this faint Gaussian nuclear component (which
accounts for 0.5$-$1.1~\% of the total galaxy light) is round
($\epsilon \sim 0.12$), we identify it as a point source (e.g.,
\citealt{2015ApJ...798...55D}). However, the disky isophotes inside
$0\farcs8$ may suggest the presence of a nuclear disk (instead of a
point source) that is poorly constrained due to the nuclear dust
spiral and PSF of the SDSS data.

To determine the uncertainties on the fit parameters, and associated
quantities (magnitudes, flux ratios and stellar masses), we ran a
series of Monte Carlo simulations.  We take into account the rms
residual scatter of the fit (Figs.~\ref{Fig1}, \ref{Fig2} and
Fig.~\ref{Fig333}), possible contamination of the surface brightness
profile due to the bright foreground star and dust spiral, as well as
errors from incorrect sky subtraction to create 200 realisations of
the galaxy's light profile. Each of these realisations was decomposed
into four components in a manner similar to the aforementioned
modelling of the actual light profile of the
galaxy. Tables~\ref{Table1} and \ref{Table2} show the $1\sigma$
uncertainties calculated using the best-fitting parameters from the
200 light profile decompositions.

\subsection{Bulge identification: Pseudo vs. classical}

This work focuses on NGC~7015's bulge. Importantly, we want to
determine whether the bulge is a `classical bulge' formed through
violent merger processes or if it is a `pseudo-bulge' built out of the
disk material via secular evolution or formed by other mechanisms.
Here, we do so by considering the S\'ersic index, ellipticity, position
angle (P.A.), $B_{4}$ and bulge-to-total flux ratio ($B/T$) of the galaxy.

Having an ellipticity $\epsilon_{\rm bul} \sim 0.19^{+0.06}_{-0.03}$,
the bulge with a low S\'ersic index $n \sim 1.5-1.9$ and a major-axis
half-light radius of $R_{\rm e} \sim 5\farcs2 \approx 1.7$ kpc
dominates the galaxy light in the inner regions ($R \la 15\arcsec$),
Figs.~\ref{Fig1}, \ref{Fig2} and Tables~\ref{Table1} and
\ref{Table2}. We note that most pseudo-bulges have S\'ersic index
$n <$ 2, while most classical bulges have $n >$ 2 (e.g.,
\citealt{2004ARA&A..42..603K, 2016ASSL..418..431K}). The outer disk
with $\epsilon_{\rm disk} \sim 0.32^{+0.04}_{-0.07}$ dominates at
$R > 15\arcsec$. Secularly-built bulges tend to have similar
ellipticities and position angles as their associated outer disks
(e.g., \citealt{2004ARA&A..42..603K, 2008AJ....136..773F}). However,
pseudo-bulges, akin to classical bulges, can be rounder than their
disks (e.g., \citealt[their Section 4]{2008AJ....136..773F};
\citealt{2015ApJ...799..226E};
\citealt{2016MNRAS.462.3800D}). \citet{2008AJ....136..773F} noted that
the formation of pseudo-bulges that are flatter than their outer disks
can be due to bar bucklings and/or unstable disks which move stars
above the plane of the disks (e.g., \citealt{1990ApJ...363..391P};
\citealt{2006ApJ...637..214M}). For NGC~7025,
$\epsilon_{\rm bul} /\epsilon_{\rm disk} \sim 0.59
^{+0.20}_{-0.16}$. The ratio of average ellipticity of the bulge to
that of the disk for 30\% of the 53 pseudo-bulges in \citet[their
Fig.\ 11 and Table~3]{2008AJ....136..773F} is
$\langle\epsilon_{\rm bul}\rangle /\langle\epsilon_{\rm disk} \rangle
\la 0.60$. For about half of their pseudo-bulges, the values of
$\langle\epsilon_{\rm bul}\rangle /\langle\epsilon_{\rm disk} \rangle$
or
$\langle\epsilon_{\rm disk}\rangle /\langle\epsilon_{\rm bul} \rangle$
are $\la 0.70$ and for 10\% of their pseudo-bulges,
$\langle\epsilon_{\rm bul}\rangle /\langle\epsilon_{\rm disk}
\rangle\sim 0.60$, similar to that of NGC~7025. Omitting the most
PSF-affected region, the position angle of the bulge of NGC~7025 is
only modestly (i.e., $5-10^{\circ}$) twisted from that of the outer
disk (Fig.~\ref{Fig1}, see also Fig.~\ref{Fig2} and
Table~\ref{Table2}). Pseudo-bulges displaying such degree of alignment
with their outer disks are presented in, e.g., \citet[their Figs.~10
and 12]{2004ARA&A..42..603K}, \citet[their Figs. 3 and
5]{2015ApJ...799..226E} and \citet[their
Fig.~4]{2016MNRAS.462.3800D}. Importantly, the outer part of
NGC~7025's bulge has the same position angle as its disk.

The quantity $B_{4}$ in the output of IRAF/{\sc ellipse} quantifies
the deviations of isophotes from pure ellipses: $B_{4}$ is
negative/positive for boxy/disky isophotes
\citep{1987MNRAS.226..747J}. Outside the PSF-affected region, the
bulge of NGC~7025 is disky inside $R \sim 5\arcsec$ turning into boxy
at $6\arcsec \la R < 10\arcsec$ before becoming disky again over
$10\arcsec \la R \la 15\arcsec$.  The $B_{4}$ values of the bulge's
disky isophotes are typically $\sim 0.010$, with $20-30$\% associated
uncertainties, in fair agreement with those of the disky isophotes
from a large portion of the galaxy's disk-dominated region
($B_{4} \sim 0.007-0.015$, see Fig.~\ref{Fig1}). Note that the values
of $|B_{4}|$ are typically $<$ 0.02 for elliptical, lenticular and
early-type spiral galaxies (e.g., \citealt[their
Fig.~4]{2012ApJ...750..121G}) but disky/boxy isophotes of galaxies
with bars, dust and/or prominent spiral structures can have $|B_{4}|$
values $>$ 0.02. Although disky isophotes in disk galaxies are
primarily due to disks, the presence of bars, rings and strong spiral
arms can result in disky isophotes. However, the latter are often
accompanied by strong isophotal twists and local extrema in
ellipticities (e.g., see \citealt{2016MNRAS.462.3800D}).

From the integration of the 1D and 2D model components, accounting for
the ellipticity of the 1D components, we measure the galaxy's
bulge-to-total and bulge-to-disk flux ratios to be
$B/T \sim 0.28-0.30$ and $B/D \sim 0.48$, respectively
(Table~\ref{Table1}).  This is in excellent agreement with
\citet{2015ApJ...799..226E} who found mean $B/T \sim 0.33$ for his
sample of pseudo-bulges (see also \citealt[their
Fig.~11]{2008AJ....136..773F}). The 1D decomposition (Fig.~\ref{Fig1})
gives a dust corrected SDSS $i$-band absolute magnitude for the bulge
$M_{i,\rm bul}\sim -22.53$ mag, while our {\sc imfit} and {\sc galfit}
2D decompositions yield $M_{i,\rm bul}\sim -22.59$ mag and
$\sim -22.55$ mag, respectively. We correct the observed
absolute bulge magnitudes for inclination and internal dust extinction
\citep[their Table 1 and Equations 1 and 2]{2008ApJ...678L.101D},
Galactic extinction and surface brightness dimming. These $i$-band
absolute magnitudes were converted into solar luminosities using an
absolute magnitude for the Sun of $M_{i,\sun} = 4.53$ AB
mag\footnote{\url{http://mips.as.arizona.edu/~cnaw/sun.html}}. To
convert the luminosities into stellar masses $M_{*,\rm bul}$, we
computed the bulge's mass-to-light ratio ($M/L_{i,\rm bul}$) using its
$g-i$ color of 1.40 mag (Fig.~\ref{Fig1}) and the color$-$($M/L$)
relation by \citet[their Table 3]{2013MNRAS.430.2715I}. This yields
$M/L_{i,\rm bul} \sim 0.63$, thus
$M_{*,\rm bul} \sim (4.21- 4.47)\times 10^{10} M_{\sun}$.

In summary, the ellipticity, position angle, $n$, $B/T$ and
predominantly disky isophotes of NGC~7025's bulge together with the
presence of dust and stellar spiral structures in the galaxy
(Fig.~\ref{Fig3}), strongly suggest a modestly flattened
pseudo-bulge. Indeed, \citet{2016ASSL..418..431K} noted that classical
bulges cannot have spiral structures.  Building on our structural
analysis, in Section~\ref{Sec3.2} we present the stellar kinematic of
the bulge.

\subsubsection{Comparison with past decomposition}\label{3.2.1}

Recently, \citet{2017A&A...604A..30N}, \citet{2018MNRAS.476.2137R} and
\citet{2018MNRAS.477..845G} fit a 2D S\'ersic bulge + exponential disk
model to the SDSS images of NGC~7025, without accounting for the
intermediate (spiral-arm) component.  As warned by \citet [see also
\citealt{2012ApJ...755..163D,2013ApJ...768...36D,2014MNRAS.444.2700D,
  2016ApJ...831..132G}]{2016MNRAS.462.3800D, 2017MNRAS.471.2321D,
  2018MNRAS.475.4670D}, neglecting to fit intermediate-scale
components (including bars, disks and spiral arms) as a separate
component can systematically bias the S\'ersic index and flux of the
inner bulge towards higher values. Indeed, \citet{2017A&A...604A..30N}
argued that their S\'ersic index for the bulge of NGC~7025
($n \sim 2.7$), the bulge's position in the Kormendy relation together
with the galaxy's concentration index and velocity dispersion gradient
all favor a merger-built `classical' bulge with $B/T \sim
0.44$. Similarly, due to the flux of the intermediate (spiral-arm)
which was attributed to the bulge, the bulge-disk decomposition by
\citet{2018MNRAS.477..845G} yielded a high S\'ersic index for bulge
($n\sim 5.1$), indicating a slow-rotating classical bulge.
\citet{2018MNRAS.476.2137R} classify NGC~7025 as an early-type
S0a. Their modeling of the bulge light distribution for this galaxy
yielded $n \sim 2.4$ and $B/T \sim 0.37$, both these figures are
larger than ours (see Tables~\ref{Table1} and \ref{Table22}).

\subsection{Spectroscopy}\label{Sec3.2}

\begin{figure*}
\begin{center}
\setlength{\tabcolsep}{0.0348in}
\begin {minipage}{180mm}
\begin{tabular}{@{}llcccc@{}}
\multicolumn{1}{c}{} \\       
\hspace*{-.150882250772599cm}   
\begin{overpic}[angle=0,scale=0.328]{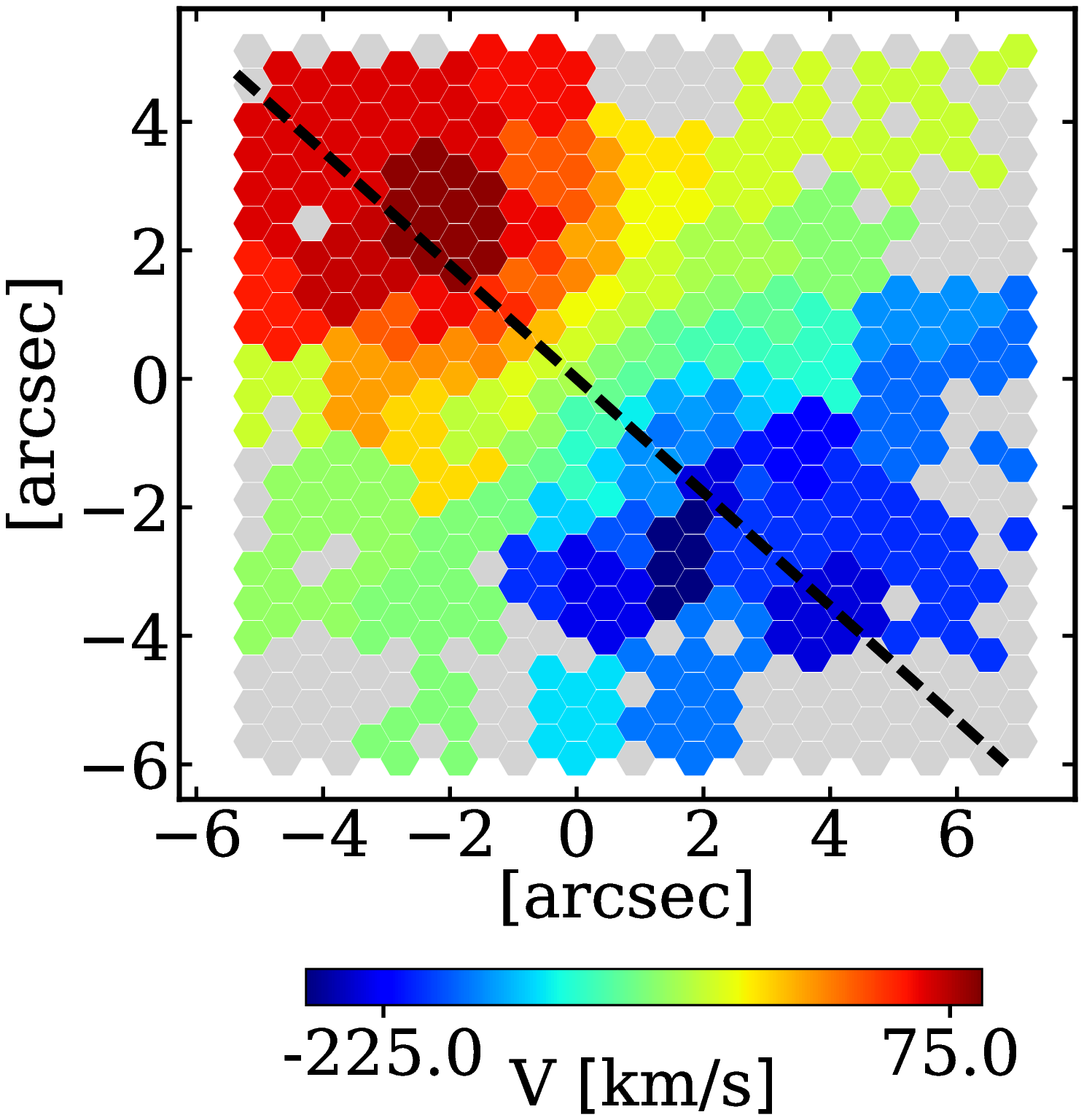}
\put(66.5,30.05){\color{black}{\large (a)}}
\put(14.5,50.05){\rotatebox{90}{\colorbox{blue}{\makebox(14,0.45){\color{white}{\tiny HR-I}}}}}
\end{overpic}
\hspace*{-1.1082250772599cm}   
\begin{overpic}[angle=0,scale=0.328]{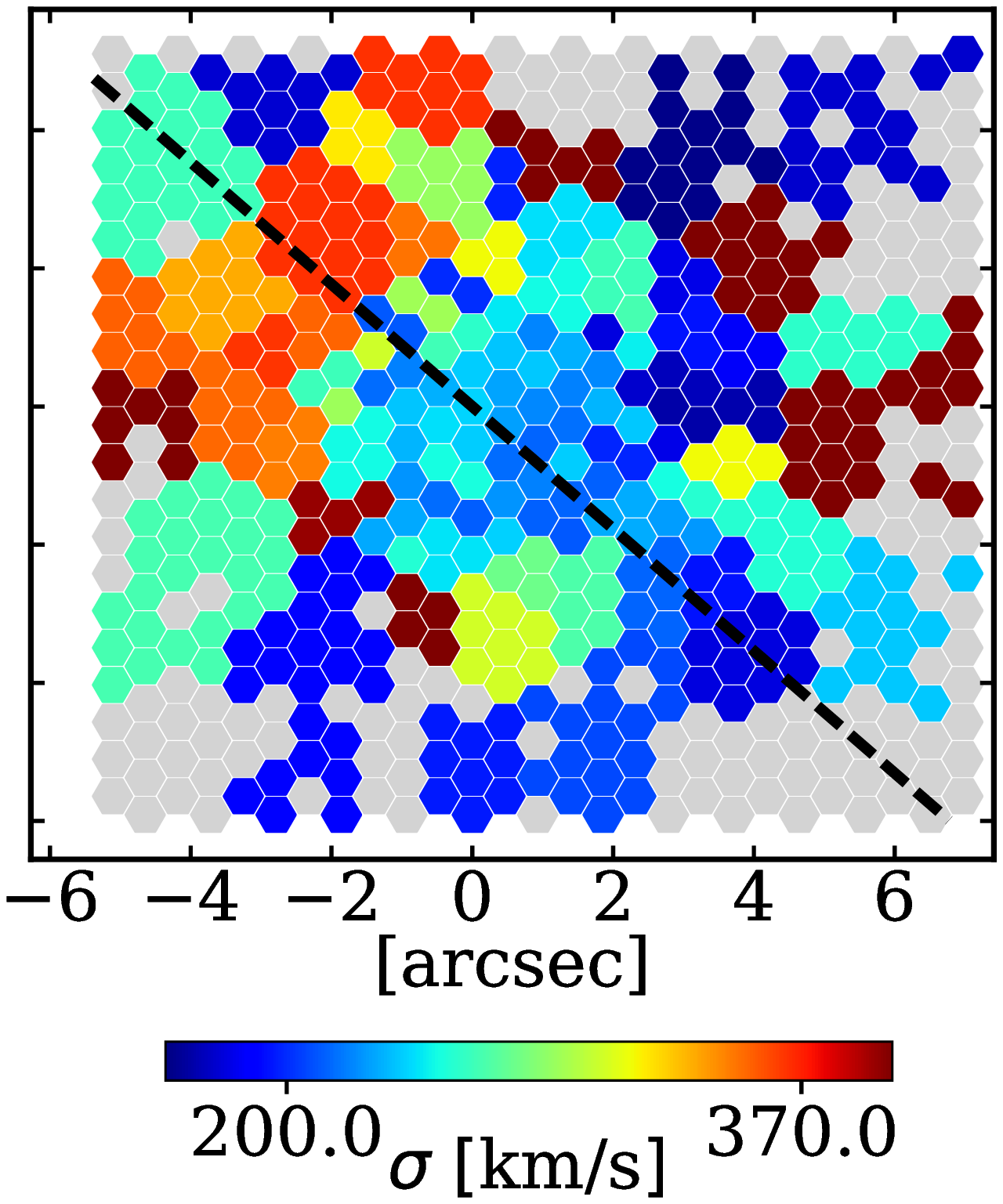}
\put(55.,30.05){\color{black}{\large(b)}}
\end{overpic}
\hspace*{-1.724050250772599cm}   
\begin{overpic}[angle=0,scale=0.328]{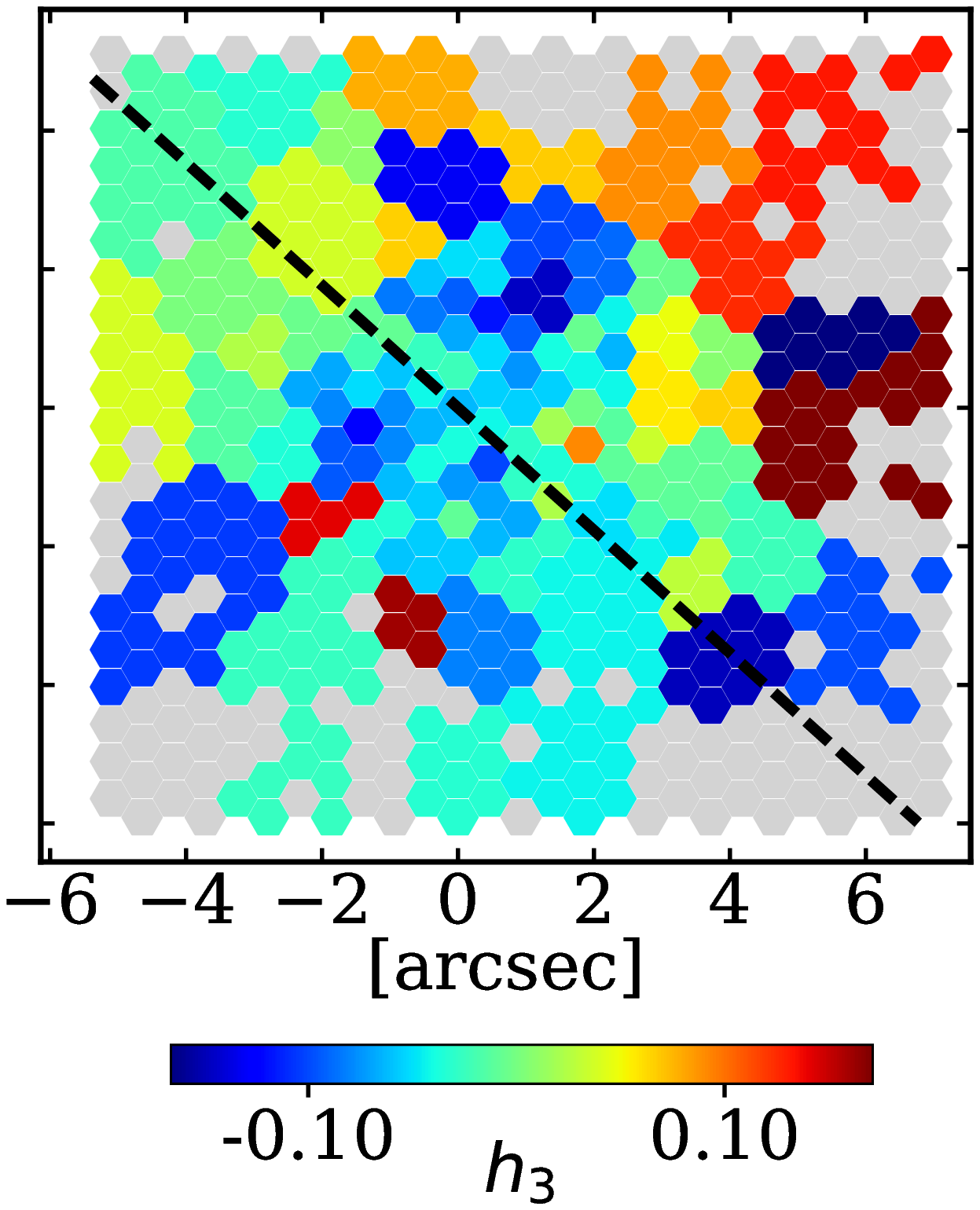} 
\put(53.,30.05){\color{black}{\large(c)}}
\end{overpic}
\hspace*{-1.882250772599cm}   
\begin{overpic}[angle=0,scale=0.328]{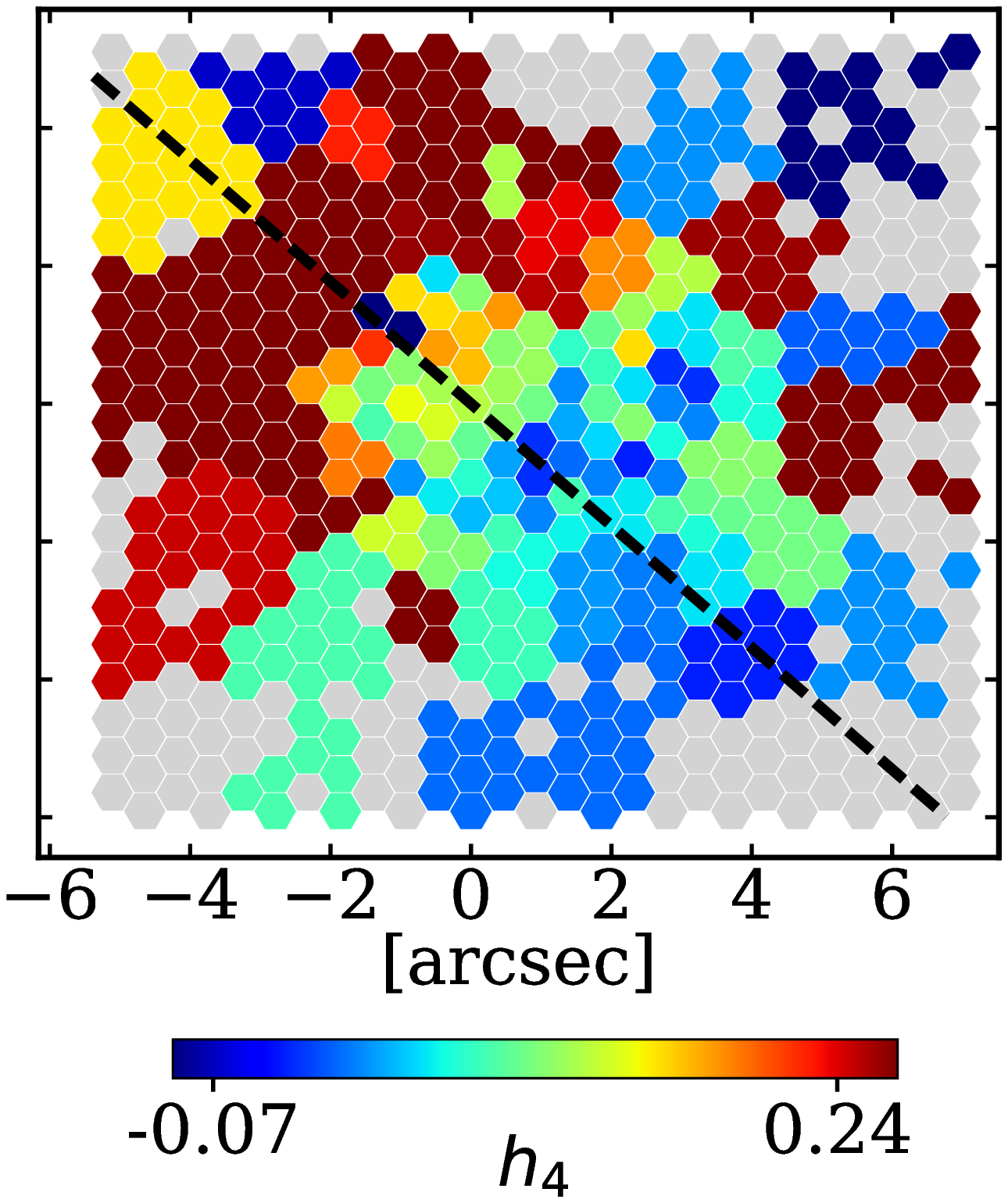}
\put(59.,30.05){\color{black}{\large (d)}}
\end{overpic}\\
\hspace*{-.150882250772599cm}   
\begin{overpic}[angle=0,scale=0.327]{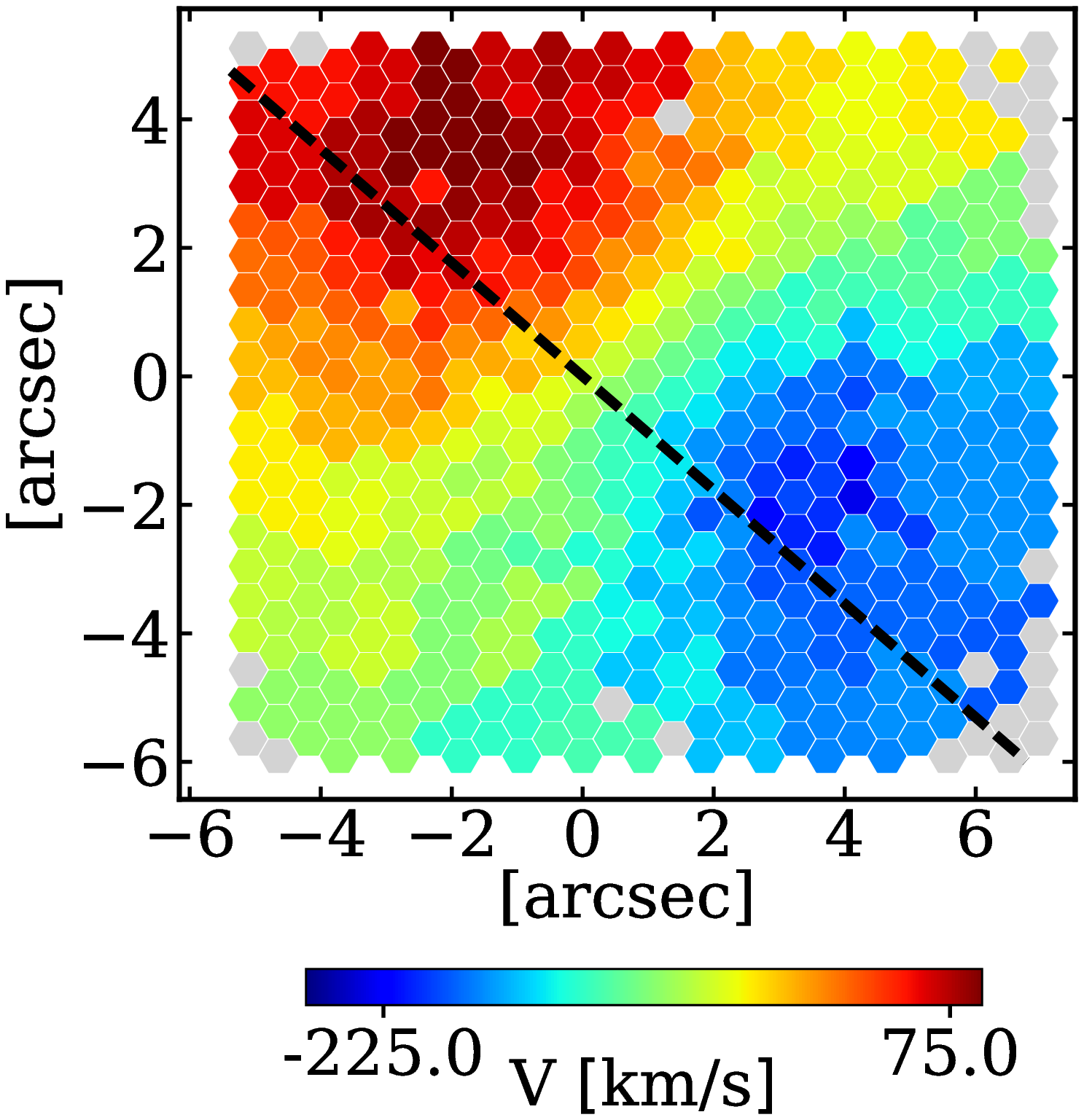}
\put(64.5,30.05){\color{black}{\large (e)}}
\put(14.5,50.05){\rotatebox{90}{\colorbox{blue}{\makebox(14,0.45){\color{white}{\tiny LR-V}}}}}
\end{overpic}
\hspace*{-1.1082250772599cm}   
\begin{overpic}[angle=0,scale=0.327]{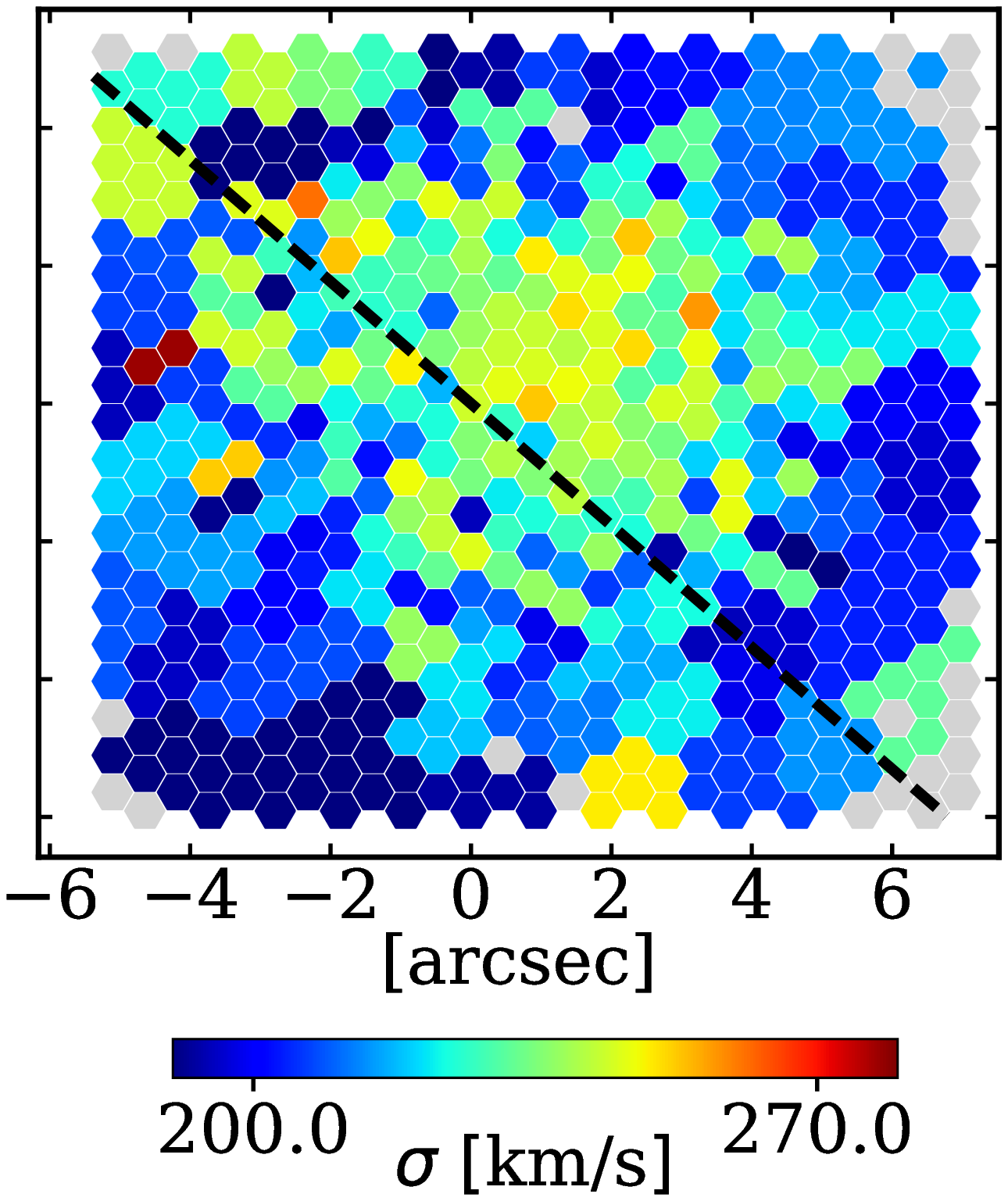}
\put(53.3,30.05){\color{black}{\large(f)}}
\end{overpic}
\hspace*{-1.924050250772599cm}   
\begin{overpic}[angle=0,scale=0.327]{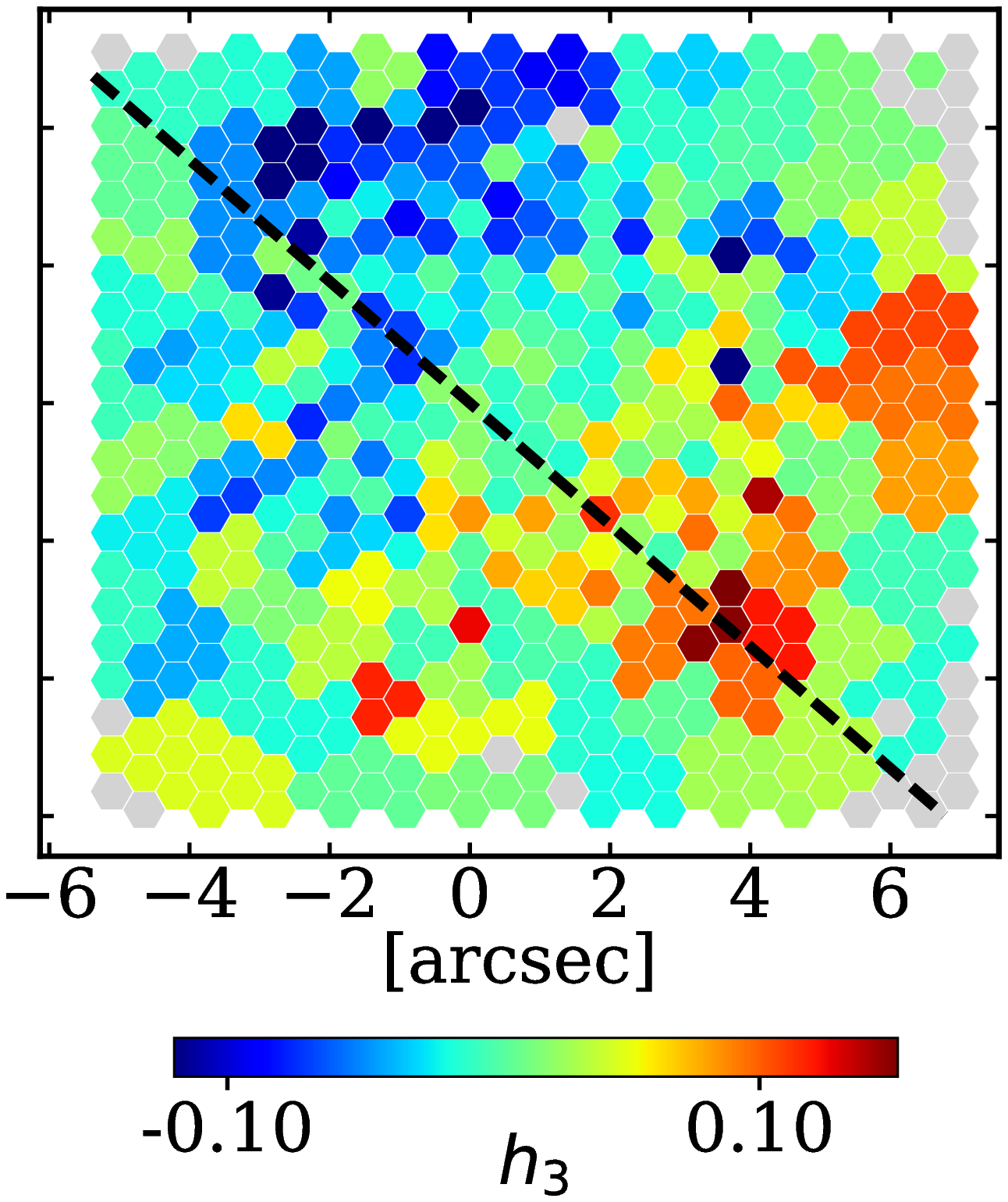} 
\put(51.,30.05){\color{black}{\large(g)}}
\end{overpic}
\hspace*{-1.8082250772599cm}   
\begin{overpic}[angle=0,scale=0.327]{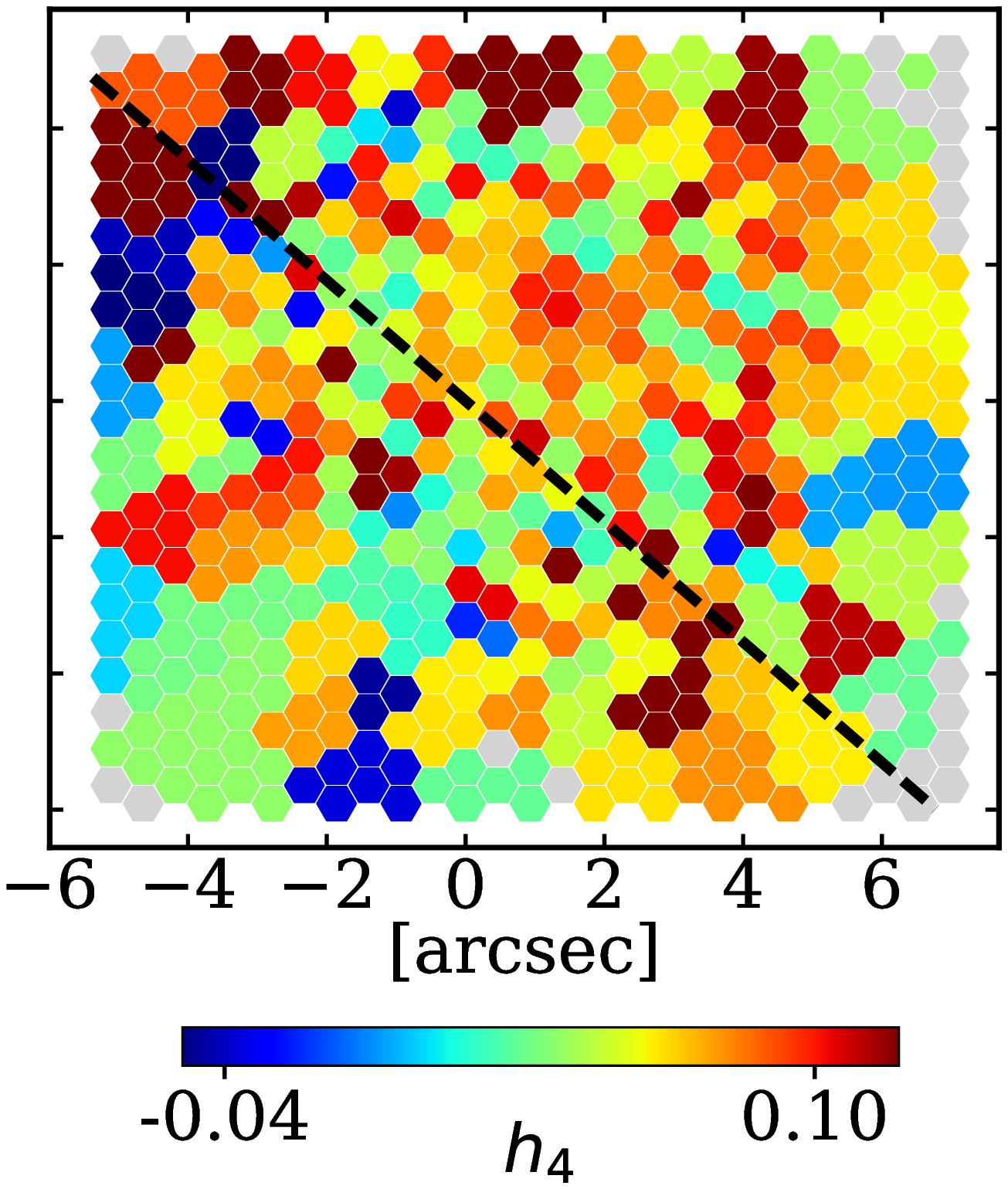}
\put(52.,30.05){\color{black}{\large (h)}}
\end{overpic}
\end{tabular} 
\caption{MEGARA stellar kinematic maps of NGC 7025 extracted from the
  HR-I spectra (top row) and LR-V spectra (bottom row). These maps
  extend to roughly the semi-major axis half-light radius of the
  pseudo-bulge ($R_{\rm e,maj} \sim 5\farcs23$). We have excluded
  fibers with $S/N < 3$, shown in light grey, and co-added the HR-I
  data into spatial bins with signal-to-noise ratios $S/N > 7$ using
  the Voronoi binning algorithm \citep{2003MNRAS.342..345C}.  The LR-V
  data were similarly co-added into spatial bins with signal-to-noise
  ratios $S/N > 15$. The panels show the velocity ($V$) determined
  with respect to NGC 7025's systemic velocity (a and e), velocity
  dispersion $\sigma$ (b and f) and the higher order Gauss-Hermite coefficients
  $h_{3}$ (c and g) and $h_{4}$ (d and h).  The dashed lines show the
  photometric major-axis of the pseudo-bulge, which was determined
  excluding the PSF-affected region (see Fig.~\ref{Fig1}). North is up
  and east is to the left. }
\end {minipage}
\label{Fig5} 
\end{center}
\end{figure*}


\subsubsection{Stellar kinematics }\label{Sec3.2.1}

\begin{figure}
\begin{center}
\begin{tabular}{@{}llcccc@{}}
\multicolumn{1}{c}{} \\       
\hspace*{-.13082250772599cm}   
\begin{overpic}[angle=0,scale=0.28]{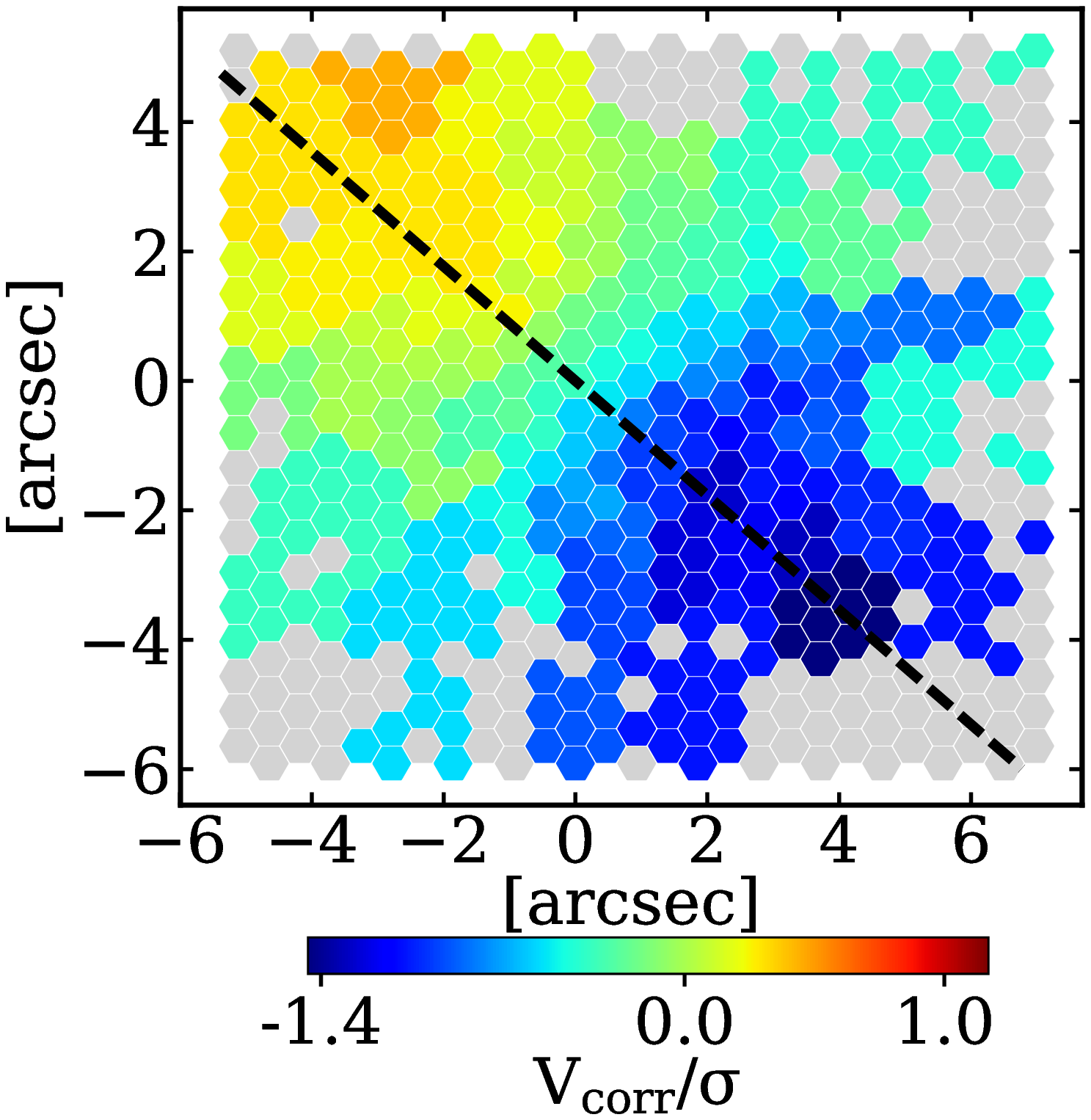}
\put(58.75,30.05){\color{black}{\large (a)}}
\put(14.26,50.05){\rotatebox{90}{\colorbox{blue}{\makebox(14,0.45){\color{white}{\tiny HR-I}}}}}
\end{overpic}
\hspace*{-.9382250772599cm}   
\begin{overpic}[angle=0,scale=0.28]{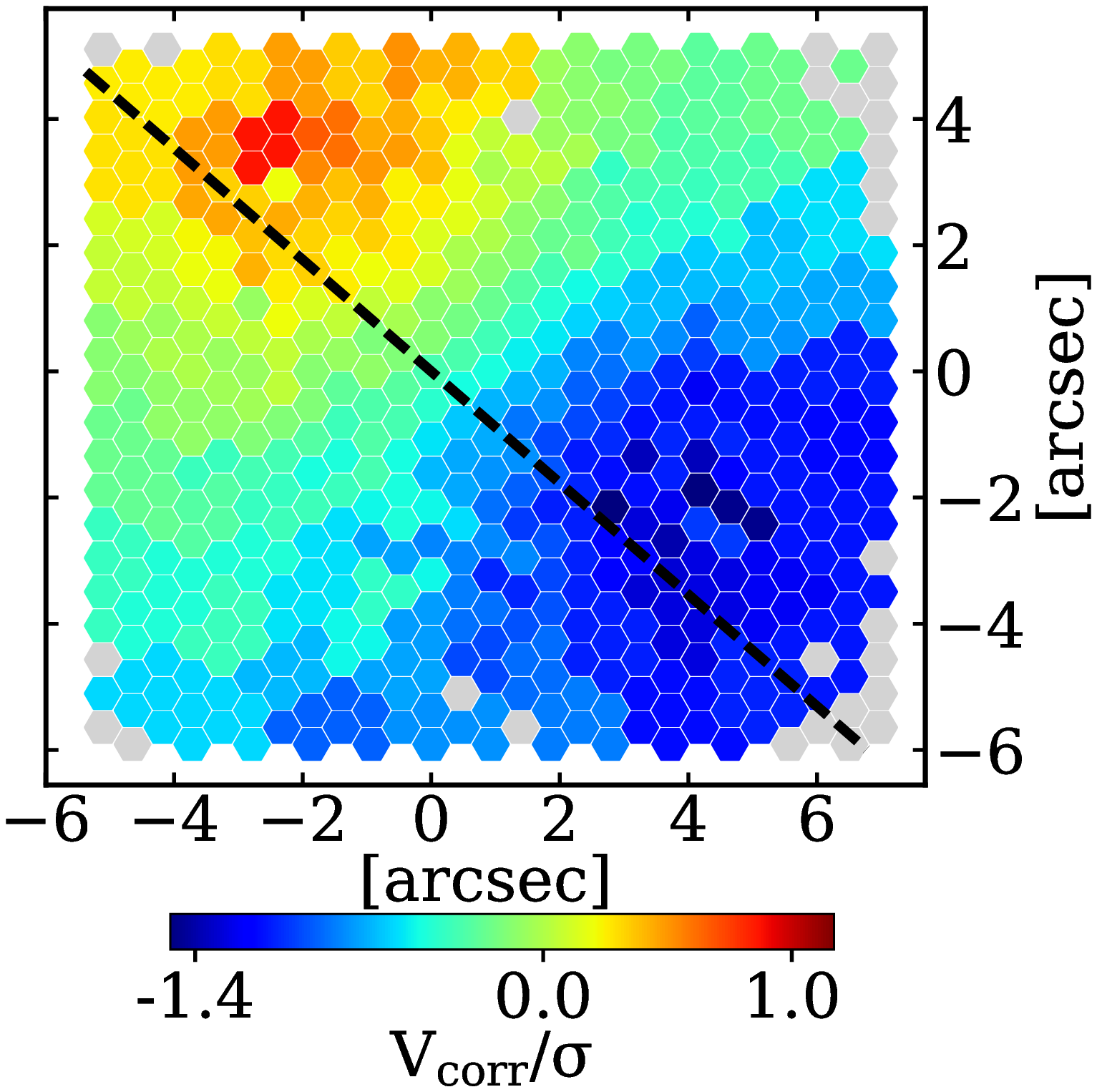}
\put(50.3,30.05){\color{black}{\large(b)}}
\put(.47625,50.05){\rotatebox{90}{\colorbox{blue}{\makebox(14,0.45){\color{white}{\tiny LR-V}}}}}
\end{overpic}
\vspace*{.0772599cm}   
\end{tabular} 
\caption{Similar to Fig.~\ref{Fig5} but here showing the ratios of the
  inclination corrected stellar rotation velocity ($V_{\rm corr}$) to
  velocity dispersion ($\sigma$) for the disky pseudo-bulge in NGC
  7025. We correct the stellar velocity for inclination ($i \sim$ 53.3
  $^{\circ})$ as
  $V_{\rm corr} = V/{ {\rm sin} (i)}$.  $|V_{\rm corr}/\sigma|$ rises
  to $\sim$1.5, indicating that the pseudo-bulge is supported by rotation.}
\label{Fig55} 
\end{center}
\end{figure}

The MEGARA HR-I and LR-V spectra for NGC~7025 have a high
signal-to-noise ratio per spaxel (S/N) of 12 $-$ 100 at the central
regions, compared to the low S/N of $\sim$1$-$5 at the outer parts of
the IFU. We excluded all fibers with S/N $<$ 3, which are
shown in light grey in Fig.~\ref{Fig5}, and spatially binned the data
using the 2D Voronoi binning technique by \citet{2003MNRAS.342..345C}.
For the HR-I spectra, we find that forcing each bin's minimum S/N
threshold to 7 results in a reliable measurement of the stellar
rotation and stellar velocity dispersion for the pseudo-bulge of
NGC~7025, with optimum spatial resolution over $R \la 7\arcsec$. For
the LR-V spectra, with higher S/N than that of the HR-I data, a
minimum S/N threshold of 15 allows a robust stellar kinematic
measurement for the pseudo-bulge.

We fit the binned MEGARA HR-I and LR-V spectra with the E-MILES and
MILES stellar templates\footnote{\url{http://miles.iac.es}},
respectively \citep{2016MNRAS.463.3409V}, to determine the stellar
kinematics (the stellar rotation $V$, stellar velocity dispersion
$\sigma$, skewness $h_{3}$ and kurtosis $h_{4}$) of the galaxy using
the penalised Pixel-Fitting code (p{\sc pxf}) by \citet[see also
\citealt{2017MNRAS.466..798C}]{2004PASP..116..138C}, Figs.~\ref{Fig4},
\ref{Fig4II} and \ref{Fig5}. These template spectra span range in age
(0.063 Gyr to 17.78 Gyr) and in metallicity (-2.32 $<$ [M/H] $<$
+0.22). We adopt a Salpeter initial mass function
\citep{1955ApJ...121..161S} for this massive spiral galaxy as
suggested by \citet{2013MNRAS.428.3183D}. The \mbox{E-MILES}/MILES
templates that we used have a spectral resolution of FWHM = 2.50
\AA~and velocity dispersions $\sigma \sim 40$ km s$^{-1}$ and $60$ km
s$^{-1}$ at the central wavelengths of the MEGARA HR-I and LR-V
spectra, respectively \citep[their Fig.\
8]{2016MNRAS.463.3409V}. These can be compared to the MEGARA spectral
resolutions of FWHM $\sim$ 0.42 \AA \ $\approx$ 15 km~s$^{-1}$ and
FWHM $\sim$ 0.95 \AA \ $\approx$ 50 km~s$^{-1}$ for the \mbox{HR-I}
and \mbox{LR-V} spectra, respectively (see \ref{Sec2.1.2}). Given
NGC~7025's bulge has broad stellar absorption features (i.e.,
$\sigma \ga 155$ km s$^{-1}$), it was not necessary to convolve the
high-resolution MEGARA spectra to degrade them to the resolution of
the \mbox{E-MILES}/MILES template spectra (Figs.~\ref{Fig4},
\ref{Fig4II} and \ref{Fig5}).

\begin{figure}
\begin{center}
\setlength{\tabcolsep}{0.0348in}
\begin{tabular}{@{}llcccc@{}}
\multicolumn{1}{c}{}       
\hspace*{-.5829152982250772599cm}     
\includegraphics[angle=0,scale=0.2650505]{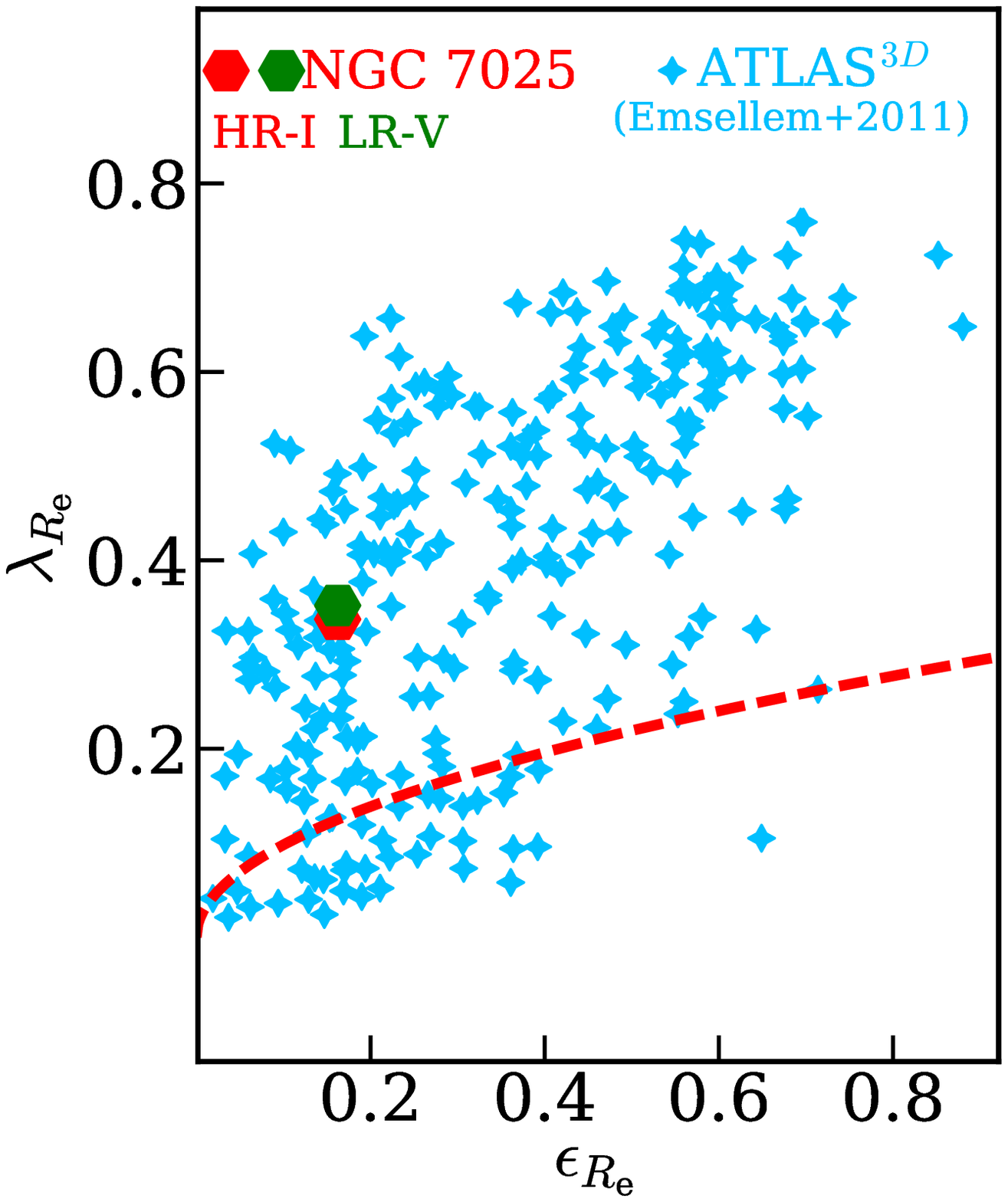}& 
\hspace*{-1.145963184240982250772599cm}     
\includegraphics[angle=0,scale=0.2650505]{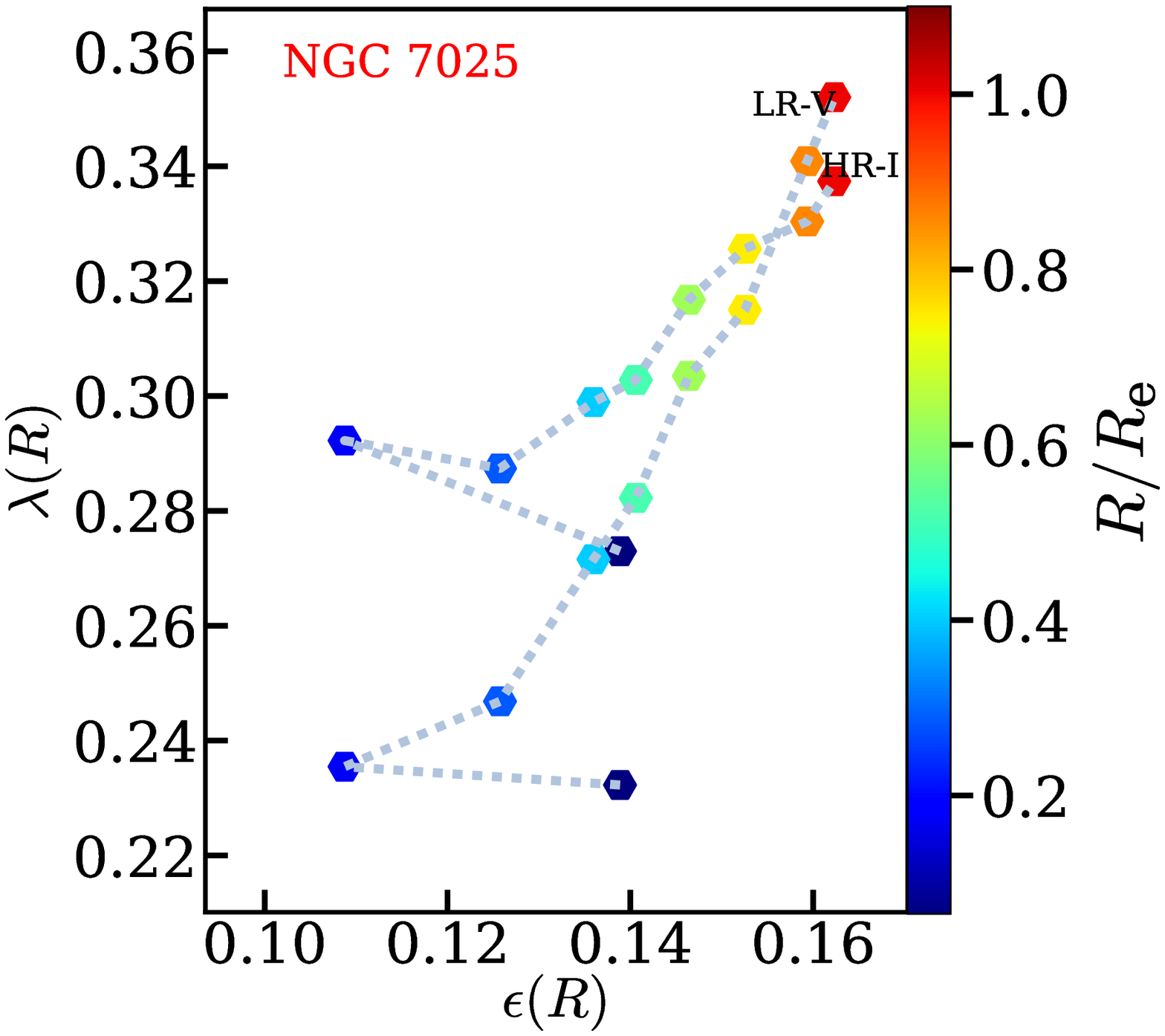}&
\hspace*{.05250772599cm}   
\end{tabular} 
\vspace*{-.40402772599cm}   
\caption{Left-hand panel: the division of galaxies into fast and slow
  rotators in the spin parameter $\lambda_{\rm e}$ versus ellipticity
  $\epsilon_{\rm e}$ diagram. Fast (slow) rotators are above (below)
  the dashed demarcation line ($\lambda_{\rm e}=$ 0.31 $\times$
  $\sqrt \epsilon_{\rm e}$) adopted by
  \citet{2011MNRAS.414..888E}. For NGC~7025, we show two $\lambda_{\rm e}$
  values derived based on the  {\sc pPXF} best fits to the galaxy's HR-I and
  LR-V spectra (Figs.~\ref{Fig4} and \ref{Fig4II}). Right-hand panel:
  trails of $\lambda(R)$ versus $\epsilon (R)$ for NGC~7025 showing an
  outwardly rising $\lambda$ and $\epsilon$ at 0.3
  $\la R/R_{\rm e} \la$ 1.0.}
\end{center}
\label{Fig6} 
\end{figure}

The high-resolution MEGARA stellar velocity $(V)$ maps of NGC~7025,
which extend to roughly the major-axis half-light radius of the
pseudo-bulge ($R_{\rm e,maj} \sim 5\farcs23$), reveal a significant
rotation around the photometric minor-axis of the galaxy, rising up to
$ |V| \sim 200$ \mbox{km s$^{-1}$} (Figs.~\ref{Fig5}a and
\ref{Fig5}e). The pseudo-bulge has a stellar velocity dispersion of
$\sigma $ $\sim 230$ km s$^{-1}$ near the center (i.e., at
$R \la 0.5 R_{\rm e}$), Figs.~\ref{Fig5}b and \ref{Fig5}f. At larger
radii ($0.5 R_{\rm e} \la R \la R_{\rm e} $), the $\sigma$ map
extracted using the \mbox{HR-I} data shows that large parts of the
pseudo-bulge display $\sigma \sim 270-350$ km s$^{-1}$ but there are
regions with $\sigma \sim 155-220$ km s$^{-1}$ (Fig.~\ref{Fig5}b). In
contrast, the $\sigma$ map based on the \mbox{LR-V} spectra shows that
the pseudo-bulge typically has $\sigma \sim 155-220$ km s$^{-1}$ at
$0.5 R_{\rm e} \la R \la R_{\rm e} $.  This discrepancy is primarily
due to the contrasting S/N ratios of the \mbox{HR-I} and \mbox{LR-V}
spectra, i.e., (S/N)$_{\rm HR-I} \sim 10-35$ versus
(S/N)$_{\rm LR-V} \sim 20-60$.  $\sigma$ are less reliably constrained
for the HR-I data with modest S/N $ \sim 10$. It is also in part
because the \mbox{HR-I} spectra is weakly sensitive to localised dust
absorption in the central regions of the galaxy, compared to the
\mbox{LR-V} spectra. These detailed MEGARA $V$ and $\sigma$ maps
together with the \citet[their Figs.~A10 and A27]{2017A&A...597A..48F}
low-resolution CALIFA $V$ and $\sigma$ maps, with larger radial extent
reaching $R \sim 30\arcsec$, reveal kinematic regularity in the
galaxy. That is, overall an outwardly increasing (declining) rotation
(stellar velocity dispersion). In agreement with our decomposition,
these large FOV kinematic maps by \citet{2017A&A...597A..48F} are such
that NGC~7025 exhibits a high degree of rotation ($ |V| > 150$ km
s$^{-1}$) at $R \ga 25\arcsec$, i.e., radii where the outer disk
dominates the pseudo-bulge.

Because the \mbox{HR-I} spectra have somewhat low S/N at
$R \ga 0.5 R_{\rm e}$, the corresponding $h_{3}$ and $h_{\rm 4}$
values are poorly constrained (Fig.~\ref{Fig5}c and Fig.~\ref{Fig5}d).
For the \mbox{LR-V} spectra, the $h_{3}$ map is anti-correlated with
the $V$ map, as expected for disky rotating systems (Fig.~\ref{Fig5},
see also \citealt{1994MNRAS.269..785B}; \citealt{2006MNRAS.372L..78G};
\citealt{2008MNRAS.390...93K}; \citealt{2017ApJ...835..104V}). The
pseudo-bulge typically has $h_{4} > 0$ at $R \la R_{\rm e}$
(Fig.~\ref{Fig5}h).


\subsubsection{$V/\sigma$}

The disky pseudo-bulge in NGC~7025 is evident from the decompositions
and isophotal analysis in Section~\ref{Sec3.1}. The correspondence
between the central structure of early-type galaxies and their
kinematics results in the disky/boxy, `fast rotator' (FR)/`slow
rotator' (SR) dichotomy (e.g., \citealt{1989A&A...217...35B};
\citealt{1991ApJ...383..112F}; \citealt{1997AJ....114.1771F}). The
former are thought to structurally and kinematically resemble the
rotating bulges in spiral galaxies.  The ratio of the stellar rotation
to velocity dispersion ($V/\sigma$) together with the ellipticity
$\epsilon$ has been used to determine if a system is supported by
rotation or anisotropic velocity dispersion (e.g.,
\citealt{1977ApJ...218L..43I, 1982ApJ...257...75K,
  1983ApJ...266...41D}). We follow the recent approach by, e.g.,
\citet{2012ApJ...754...67F} and \citet{2015ApJ...799..226E}, first
correct $V$ for inclination ($i$) as $V_{\rm corr} = V/{\rm sin} (i)$
and then determine $V_{\rm corr}/\sigma$ for NGC~7025 with
$i \sim53.3$  $^{\circ}$ 
(HyperLeda\footnote{http://leda.univ-lyon1.fr/search.html},
\citealt{2003A&A...412...45P}). We adopt a kinematic classification
such that rotation-supported systems have regions with
$V_{\rm corr}/\sigma \ge 1$, while for pressure-supported systems
$V_{\rm corr}/\sigma < 1$ (e.g., \citealt{2012ApJ...754...67F,
  2015ApJ...799..226E}). Fig.~\ref{Fig55} shows the HR-I and LR-V
$V_{\rm corr}/\sigma$ maps of NGC 7025's pseudo-bulge. These maps show
that $|V_{\rm corr}/\sigma|$ climbs to $\sim$1.5, revealing that the
pseudo-bulge is mainly supported by rotation.

\subsubsection{ Specific angular momentum $\lambda_{R}$}

Using the SAURON IFU data for a sample of
48 early-type galaxies, \citet{2007MNRAS.379..401E} advocated the
parameter $\lambda_{R}$ for discriminating between SRs and FRs in
their sample. This (spin) parameter $\lambda_{R}$, a proxy for the
stellar angular momentum per unit mass, is defined as

 \begin{equation}
\lambda_{R_{max}} =
\frac{\sum_{i=1}^{N_{max}} F_{i}R_{i}|V_{i}|}{\sum_{i=1}^{N_{max}}
  F_{i}R_{i}\sqrt{V_{i}^{2}+\sigma_{i}^{2}}}, 
\label{Eq1}
 \end{equation}
 where $F_{i}$, $R_{i}$, $V_{i}$ and $\sigma_{i}$ are the flux,
 circular radius, velocity and velocity dispersion of the ith spatial
 bin, respectively. Having measured the half-light spin parameter
 $\lambda_{\rm e} = \lambda (R_{\rm e} = 4\farcs68)$ and average
 ellipticity of the galaxy within $R_{\rm e}$ ($\epsilon_{\rm e}$), in
 Fig.~\ref{Fig6} (left) we place NGC~7025 in the
 ($\lambda_{\rm e} -\epsilon_{\rm e}$) diagram and compare it with the
 ATLAS$^{\rm 3D}$ sample of 260 early-type galaxies
 \citep{2011MNRAS.414..888E}.  Fast rotators are defined by
 \citet{2011MNRAS.414..888E} as those galaxies with
 $\lambda_{\rm e} > 0.31\times \sqrt \epsilon_{\rm e} $ and thus they
 occupy the region above the dashed boundary line in Fig.~\ref{Fig6}
 (left), while those with
 $\lambda_{\rm e} < 0.31\times \sqrt \epsilon_{\rm e}$ which lie below
 this boundary line are slow rotators. We note that there are
 `intermediate-type' galaxies with
 $\lambda_{\rm e} = 0.31\times \sqrt \epsilon_{\rm
   e}$. Fig.~\ref{Fig6} clearly reveals that NGC~7025's pseudo-bulge is
 a fast rotator.
  
 In addition, given the high-resolution data afforded by MEGARA, we
 attempt to characterise the rotation of the pseudo-bulge as a
 function of radius.  Fig.~\ref{Fig6} (right) plots spin tracks for
 the \mbox{HR-I} and \mbox{LR-V} data created by measuring
 $\lambda (R)$ and $ \epsilon (R)$ within 9 different circular
 apertures with radii $R $= (0.07, 0.18, 0.29, 0.41, 0.53, 0.64, 0.76,
 0.87 and 1.00)$\times$$R_{\rm
   e}$. These tracks reveal that the pseudo-bulge, which has an
 outwardly rising $\lambda (R)$ and $ \epsilon (R)$ for 0.3
 $\la$ $R/R_{\rm e}$
 $\la$ 1.0, is consistent with being a fast rotator for all the
 apertures we considered. It is worth noting that the spin tracks by
 \citet[their Fig.~9]{2017ApJ...840...68G} differ from ours. Instead
 of using circular apertures as done here, they advocated determining
 $\lambda
 (R)$ within elliptical annuli to better represent the rotation
 profiles for their galaxies with intermediate-scale disks (see also
 \citealt{2017MNRAS.470.1321B}). Given that we are interested in
 determining the radial $\lambda
 (R)$ profiles only for {\mbox NGC 7025's} pseudo-bulge component,
 $\lambda$ values measured using circular apertures are more
 meaningful than those determined within elliptical annuli.

%
%

\section{Discussion}\label{Sec4} 

\subsection{The nature of the disky bulge in NGC 7025}

High-resolution MEGARA spectroscopy of the early-type spiral (Sa)
galaxy NGC~7025, obtained as part of the commissioning run of the
MEGARA instrument, reveals a disky bulge that is supported by
rotation. As noted in the preceding sections, this bulge is well
fitted by a S\'ersic model with a low S\'ersic index ($n \sim$
1.5$-$1.9).  Its properties are generally consistent with a
secular-driven formation scenario (e.g.,
\citealt{1982ApJ...257...75K}, \citealt{1993IAUS..153..387P};
\citealt{1993IAUS..153..209K}; \citealt{1996ApJ...457L..73C};
\citealt{1997AJ....114.2366C}; \citealt{2005MNRAS.358.1477A};
\citealt{2007MNRAS.381..401L}; \citealt{2008AJ....136..773F}), as such
favoring a pseudo-bulge interpretation. In contrast,
\citet{2017A&A...604A..30N} and \citet{2018MNRAS.477..845G} advocated a
classical bulge for the galaxy. The 2D decompositions of the SDSS
images of NGC 7025 by \citet{2017A&A...604A..30N} and
\citet{2018MNRAS.477..845G} were performed without accounting for the
galaxy intermediate (spiral-arm) component (see also
\citealt{2018MNRAS.476.2137R}), this biased their S\'ersic index and
flux for the bulge component towards higher values, which are expected
for classical bulges (see Section~\ref{3.2.1}).

As mentioned in the Introduction, in the secular evolution picture,
pseudo-bulges are thought to have formed (slowly) via inward
funnelling of disk materials facilitated by non-axisymmetric features,
such as bars, ovals, lenses or spiral arms (e.g.,
\citealt{1990ApJ...363..391P}; \citealt{1993IAUS..153..209K};
\citealt{2004ARA&A..42..603K}; \citealt{2007MNRAS.381..401L};
\citealt{2008AJ....136..773F}; \citealt{2008MNRAS.388.1708G};
\citealt{2010MNRAS.405.1089L}; \citealt{2014A&A...572A..25M};
\citealt{2015ApJ...799..226E}; \citealt{2016MNRAS.459.4109T}).
Therefore, the prediction is that there is a coupling between the
pseudo-bulges and their surrounding large-scale disks as revealed by,
for example, the nearly constant bulge-to-disk size ratios
($R_{\rm e}/h$) of $\sim$ 0.20 and 0.24 for late- and early-type
spirals, respectively (\citealt{2003ApJ...582..689M}, see also
\citealt{1996ApJ...457L..73C}; \citealt{2001AJ....121..820G};
\citealt{2008AJ....136..773F}; \citealt{2008MNRAS.388.1708G};
\citealt{2009MNRAS.393.1531G}).  \citet{2010MNRAS.405.1089L} found
median $R_{\rm e}/h$ values of $0.15$ and 0.20 for their S0/a and S0
galaxies.  For NGC 7025, we find $R_{\rm e}/h \sim 0.18-0.20$
(Table~\ref{Table1}). Given that \citet{2003ApJ...582..689M} did not
account for spiral arms in their light profile decompositions, our
$R_{\rm e}/h$ value for NGC 7025 is generally in good agreement with
those of the early-type spiral and S0 galaxies
\citep{2003ApJ...582..689M, 2010MNRAS.405.1089L}.

NGC~7025's pseudo-bulge fulfils three pseudo-bulge identification
criteria by \citet{2004ARA&A..42..603K}: \\
\begin{itemize}
 \item Having a rotationally
supported dynamics (Figs.~\ref{Fig5}, \ref{Fig55} and \ref{Fig6}).

 \item Having a
nearly exponential light profile (Figs.~\ref{Fig1}, \ref{Fig2} and
Tables~\ref{Table1} and \ref{Table22}) and the presence of dust and
stellar spiral structures which extend all the way into the galaxy
center (Figs.~\ref{Fig3} and \ref{Fig333}). 

\item Furthermore, our $n$
values for pseudo-bulge agree with the \citet{2008AJ....136..773F}
pseudo-bulge ($n \la 2$) versus classical bulge ($n \ga 2$) divide
(see also \citealt{2017A&A...604A..30N} who advocated a similar $n$
based pseudo-bulge/classical bulge dichotomy but using a threshold
$n =1.5$ instead of 2). However, we refer the reader to a detailed
discussion by \citet[see his Section 4.3]{2013pss6.book...91G} who
highlighted the danger of using such bulge classification criteria alone.
\end{itemize}

Our pseudo-bulge-to-total flux ratio for NGC~7025 ($B/T \sim 0.28-30$)
is in excellent agreement with \citet{2015ApJ...799..226E} who
reported mean $B/T$ of $\sim 0.33$ for his sample of pseudo-bulges
(see also \citealt[their Fig. 11]{2008AJ....136..773F}) and is akin to
the $B/T$ values of $\sim 0.25-0.35$ typically reported for other
spiral and lenticular galaxies (see e.g.,
\citealt{2008MNRAS.388.1708G}; \citealt{2008AJ....136..773F};
\citealt{2009MNRAS.393.1531G}; \citealt{2009ApJ...696..411W};
\citealt{2010MNRAS.405.1089L}; \citealt[references
therein]{2016MNRAS.462.3800D}).

Pseudo-bulges and their surrounding disks are expected to have similar
colors, relatively bluer than those of classical bulges built in
violent merging events (e.g., \citealt{1996AJ....111.2238P};
\citealt{1997AJ....114.2366C}, \citealt{2008AJ....136..773F};
\citealt{2007ApJ...657L..85D}; \citealt{2009MNRAS.393.1531G};
\citealt{2009ApJ...699..105C}). Excluding the most PSF affected region
($R\la 1\arcsec$) and going out from the center, the $g-i$ colour
profile of NGC~7025 plotted in Fig.~\ref{Fig1} becomes gradually
bluer, it then levels at intermediate radii
($\sim 10\arcsec \la R \la 40\arcsec$) before turning redder at large
radii where the disk dominates. Not only does the galaxy color profile
agree with the secular evolution scenario but it also resembles the
`$\bigcup$-shaped' color profile of `Type II' galaxies \citep[their
Fig.~1]{2008ApJ...683L.103B}, having truncated exponential profiles
believed to arise from disk instabilities driven by bars (e.g.,
\citealt{2006A&A...454..759P}; \citealt{2016A&A...591L...7B};
\citealt{2017ApJ...848...87C};
\citealt{2018ApJS..234...18B}). However, the $g-i$ color of NGC~7025
($\sim 1.49 \pm 0.11$, see Fig~\ref{Fig3}) is markedly redder than the
typical $g-i$ = 1.06 $\pm$ 0.17 color reported for Sa galaxies
\citep[their Table~1]{2001AJ....122.1238S}. We have converted their
$g^{*}-i^{*}$ colors into $g-i$ using the SDSS DR7 photometric
equations\footnote{http://classic.sdss.org/dr7/algorithms/fluxcal.html}. The
presence of dust may partly explain the redder color of the galaxy.

\subsection{Formation of NGC 7025's pseudo-bulge}

Within the secular evolution formation scenario, massive pseudo-bulges
are closely linked to strong bars, owing to the high efficiency of
such bars at funnelling disk materials towards the inner regions of
galaxies (e.g., \citealt{2004ARA&A..42..603K};
\citealt{2008MNRAS.390..881D}; \citealt{2013ApJ...779..162C}). Although
NGC 7025 does not contain a bar or an oval structure, its compact
($R_{\rm e} \sim 1.48 - 2.06$ kpc) pseudo-bulge is 
massive, i.e., a stellar mass
$M_{*,\rm bulge} \sim (4.21-4.47) \times 10^{10} M_{\sun}$, only a
factor of 1.66 $-$ 3.33 less massive than those of the 21 local
compact ($R_{\rm e} \la 2$ kpc) and massive
($0.7 \times 10^{11} M_{\sun} < M_{*} < 1.4 \times 10^{11} M_{\sun}$)
bulges identified by \citet{2015ApJ...804...32G}. The latter bulges
were shown to have similar physical properties as the compact, massive
high-redshift galaxies found at $z \sim 1.5-2$
\citep{2013pss6.book...91G, 2013ApJ...768...36D, 2014ASPC..480...75D,
  {2015ApJ...804...32G}, 2016MNRAS.457.1916D}. We tentatively propose
that a secular process involving the tightly wound stellar spiral arms
of NGC~7025 may drive gas and stars out of the disk into the inner
regions of the galaxy, building up its massive pseudo-bulge (e.g.,
\citealt{1997AJ....114.2366C}; \citealt{2004ARA&A..42..603K};
\citealt{2007MNRAS.381..401L}). Indeed, NGC~7025 is the only unbarred
isolated galaxy from a sample of 49 CALIFA galaxies by
\citet{2015MNRAS.451.4397H} which exhibits non-circular, bar-like
flows. Simulations have shown bar dissolution caused by central mass
concentrations (e.g., \citealt{1996ApJ...462..114N};
\citealt{2004ApJ...604..614S}; \citealt{2005MNRAS.363..496A}) but this
process is not well understood. Given the low concentration of the
light distribution of NGC~7025's pseudo-bulge as revealed by the low
$n$ value, it seems unlikely that the galaxy had a bar in the past
that was destroyed by the above process.

\subsection{Alternative pseudo-bulge formation scenarios}

Alternative pseudo-bulge formation scenarios have been discussed in
the literature, for example, simulations have shown pseudo-bulges
built via high-redshift starbursts with minor contributions from
secular evolution \citep{2013MNRAS.428..718O} and through gas-rich
minor or/and major galaxy merger events (e.g.,
\citealt{2005ApJ...622L...9S}; \citealt{2011A&A...533A.104E};
\citealt{2013ApJ...772...36G}; \citealt{2015A&A...579L...2Q};
\citealt{2016ApJ...821...90A}; \citealt{2016MNRAS.459.4109T};
\citealt{2018MNRAS.473.2521S}). In the former scenario, the starburst
built pseudo-bulges form before their surrounding disks and they are
already in place at redshift of $z \sim 2-3$. This is similar to the
high-redshift clump pseudo-bulge formation scenario by
\citet{2012MNRAS.422.1902I} but both these high-redshift scenarios are
inconsistent with (i) the simulations by \citet{2008ApJ...688...67E}
which revealed that the coalescence of \mbox{high-redshift} clumps due
to disk instabilities leads to the formation of bulges that resemble
classical bulges and (ii) the color profile of NGC~7025 mentioned
above (see Fig.~\ref{Fig1}).  For the merger driven scenarios, the
pseudo-bulges tend to have young stellar populations (except for those
by \citealt{2013ApJ...772...36G}) and they are formed prior to their
surrounding disk. However, NGC~7025 is an isolated spiral galaxy, its
low-density environment does not favour the merger-driven
scenario. Indeed, $\sim 88\%$ of the sample of isolated galaxies
studied by \citet{2013MNRAS.433.1479H} have experienced at most one
minor merger over their lifetime (see aslo
\citealt{2008MNRAS.390..881D}). Also, \citet{2012ApJ...756...26M} reported
that most of their isolated spiral galaxies with $B/T < 0.3$
experienced only minor mergers since $z \sim 2$. Using theoretical
galaxy formation models, \citet[their Figure 9]{2016MNRAS.459.4109T}
also reported that bulges built through secular evolution dominate the
galaxy mass distribution at intermediate masses,
$10 \la {\rm log} ~(M_{*}/M_{\sun}) \la 11$, in contrast merger built
bulges dominate at the lower- and higher-mass ends (see also
\citealt{2012ApJ...756...26M}).

Furthermore, it has been shown that galaxy mergers or accretion of
external gas can produce kinematically distinct (counter-rotating)
stellar components in galaxies (e.g., \citealt{1984ApJ...287..577K},
\citealt{1990ApJ...361..381B}; \citealt{1992ApJ...401L..79B};
\citealt{2001AJ....121..140K}; \citealt{2011MNRAS.414.2923K};
\citealt[see their Section
4.3.2]{2018MNRAS.475.4670D}). \citet{2011MNRAS.414.2923K} identified
such kinematic features in roughly $26\%$ of their ATLAS$^{\rm 3D}$
sample of 260 early-type galaxies. The absence of any significant
peculiarity in the stellar kinematics of NGC~7025 (Fig.~\ref{Fig5} and
\citealt{2017A&A...597A..48F}) provides further support for the
secular evolution scenario but this is not to say that merger events
cannot build bulges with featureless kinematics.

\section{Conclusions}\label{Sec4} 

This paper presents the first data and scientific results of MEGARA, a
high-resolution IFU and MO spectrograph installed on the GTC.  We
carry out MEGARA IFU observations of the Sa galaxy {\mbox NGC
  7025}. The results from these observations show that MEGARA is able
to deliver high-resolution IFU data ideal for studying the central
kinematic properties of nearby galaxies. We also performed detailed 1D
and 2D decompositions of this galaxy's SDSS $i$-band data into a
Gaussian ($n=0.5$) nuclear component, a S\'ersic bulge, a S\'ersic
intermediate-scale spiral-arm component and an outer exponential
($n=1$) disk. The main results
from this work are as follows.\\

1) We find that NGC 7025's bulge has a S\'ersic index
$n \sim 1.80 \pm 0.24$, half-light radius $R_{\rm e} \sim 1.70 \pm 0.43$
kpc, stellar mass $M_{*} \sim (4.34 \pm 1.70) \times10^{10} M_{\sun}$
and bulge-to-total flux ratio $B/T \sim 0.30$.

2) We have presented the spins ($\lambda$) and ellipticities
($\epsilon$) of the bulge enclosed within nine different circular
apertures with radii $R \le R_{\rm e}$, revealing that the bulge,
which exhibits a spin track of an outwardly rising $\lambda$ and
$\epsilon$, is a fast rotator for all the apertures
considered. Correcting the bulge's stellar velocity $V$ for
inclination, we constructed $V_{\rm corr}/\sigma$ maps which show that
the bulge is mainly supported by rotation.

3) The photometric and kinematic findings given above strongly favor a
pseudo-bulge in NGC 7025, in contrast to past works which advocated a
classical bulge interpretation.

4) Our results for NGC 7025 are broadly consistent with the secular
evolution model of pseudo-bulge formation. The galaxy may have had its
disky pseudo-bulge formed naturally from the outer disk slowly driven
by the tightly wound stellar spiral arms. Examining the stellar
kinematics and population together with a careful analysis of
high-resolution {\it HST} images for a large number of unbarred
late-type galaxies is desirable to check how frequent massive
($M_{*} \sim 5 \times 10^{10} M_{\sun}$) pseudo-bulges can be built up
by secular evolution with spiral arms as the driving agents.

\section{ACKNOWLEDGMENTS}

We thank the referee for their careful reading of the paper and many
suggestions that improved the paper.  BTD acknowledges support from a
Spanish postdoctoral fellowship `Ayudas 1265 para la atracci\'on del
talento investigador. Modalidad 2: j\'ovenes investigadores.' funded
by Comunidad de Madrid under grant number 2016-T2/TIC-2039. We
acknowledge financial support from the Spanish Ministry of Economy and
Competitiveness (MINECO) under grant number AYA2016-75808-R, which is
partly funded by the European Regional Development Fund (ERDF).

Funding for the Sloan Digital Sky Survey IV has been provided by the
Alfred P. Sloan Foundation, the U.S. Department of Energy Office of
Science, and the Participating Institutions. SDSS-IV acknowledges
support and resources from the Center for High-Performance Computing
at the University of Utah. The SDSS web site is www.sdss.org.

SDSS-IV is managed by the Astrophysical Research Consortium for the
Participating Institutions of the SDSS Collaboration including the
Brazilian Participation Group, the Carnegie Institution for Science,
Carnegie Mellon University, the Chilean Participation Group, the
French Participation Group, Harvard-Smithsonian Center for
Astrophysics, Instituto de Astrof\'isica de Canarias, The Johns
Hopkins University, Kavli Institute for the Physics and Mathematics of
the Universe (IPMU) / University of Tokyo, Lawrence Berkeley National
Laboratory, Leibniz Institut f\"ur Astrophysik Potsdam (AIP),
Max-Planck-Institut f\"ur Astronomie (MPIA Heidelberg),
Max-Planck-Institut f\"ur Astrophysik (MPA Garching),
Max-Planck-Institut f\"ur Extraterrestrische Physik (MPE), National
Astronomical Observatories of China, New Mexico State University, New
York University, University of Notre Dame, Observat\'ario Nacional /
MCTI, The Ohio State University, Pennsylvania State University,
Shanghai Astronomical Observatory, United Kingdom Participation Group,
Universidad Nacional Aut\'onoma de M\'exico, University of Arizona,
University of Colorado Boulder, University of Oxford, University of
Portsmouth, University of Utah, University of Virginia, University of
Washington, University of Wisconsin, Vanderbilt University, and Yale
University.

\bibliographystyle{apj}
\bibliography{Bil_Paps_biblo.bib}

\label{lastpage}
\end{document}